\documentclass[english]{iopart}

\usepackage{graphicx}
\usepackage{url}
\newcounter{fig}
\begin{document}

\title[High order Fuchsian ODE for $\tilde{\chi}^{(5)}$ ]{\Large
High order Fuchsian equations for the square lattice
Ising model: $\tilde{\chi}^{(5)}$ }
\vskip .3cm 

\author{
A. Bostan$^\P$,
S. Boukraa$^\dag$, A. J. Guttmann$^\ddag$, S. Hassani$^\S$, I.~Jensen$^\ddag$,
J.-M. Maillard$^{||}$ and N. Zenine$^\S$}
\address{$^\P$ \ INRIA Paris-Rocquencourt, 
Domaine de Voluceau, B.P. 105
78153 Le Chesnay Cedex, France} 
\address{\dag LPTHIRM and D\'epartement d'A{\'e}ronautique,
 Universit\'e de Blida, Algeria}
\address{$\ddag$ ARC Centre of Excellence for 
Mathematics and Statistics of Complex Systems 
Department of Mathematics and Statistics,
The University of Melbourne, Victoria 3010, Australia}
\address{\S  Centre de Recherche Nucl\'eaire d'Alger, 
2 Bd. Frantz Fanon, BP 399, 16000 Alger, Algeria}
\address{$||$ LPTMC, UMR 7600 CNRS, 
Universit\'e de Paris, Tour 24,
 4\`eme \'etage, case 121, 
 4 Place Jussieu, 75252 Paris Cedex 05, France} 
\ead{alin.bostan@inria.fr, maillard@lptmc.jussieu.fr, maillard@lptl.jussieu.fr, tonyg@ms.unimelb.edu.au, 
 I.Jensen@ms.unimelb.edu.au, njzenine@yahoo.com, boukraa@mail.univ-blida.dz}

\begin{abstract}
We consider the  Fuchsian linear differential equation obtained (modulo a prime)
for $\tilde{\chi}^{(5)}$, the five-particle contribution to the susceptibility
of the square lattice Ising model. 
We show that one can understand the factorization of the corresponding
linear differential operator from calculations using just a single prime.
A particular linear combination of $\tilde{\chi}^{(1)}$ 
and $\tilde{\chi}^{(3)}$ can be removed
from $\tilde{\chi}^{(5)}$ and the resulting series 
is annihilated by a high order
globally nilpotent linear ODE. The corresponding (minimal order) linear 
differential operator, of order 29, splits into factors of small orders. 
A fifth order linear differential operator occurs as the left-most factor
of the ``depleted"  differential operator and it
is shown to be equivalent to the symmetric
fourth power of $L_E$, the linear differential operator corresponding to
the elliptic integral $E$.
This result generalizes  what we have found for the lower order terms
$\tilde{\chi}^{(3)}$ and $\tilde{\chi}^{(4)}$. We 
conjecture that a linear differential
operator equivalent to a symmetric  $(n-1)$-th power of $L_E$ 
occurs as a left-most factor in the minimal order linear 
differential operators for all $\tilde{\chi}^{(n)}$'s.
\end{abstract}

\vskip .5cm

\noindent {\bf PACS}: 05.50.+q, 05.10.-a, 02.30.Hq, 02.30.Gp, 02.40.Xx

\noindent {\bf AMS Classification scheme numbers}: 34M55, 
47E05, 81Qxx, 32G34, 34Lxx, 34Mxx, 14Kxx 

\vskip .5cm
 {\bf Key-words}:  Susceptibility of the Ising model, 
formal power series, long series expansions,
Fuchsian linear differential equations, 
globally nilpotent linear differential
operators, indicial equation, singular behavior, critical exponents,
apparent singularities, 
formal modular calculations,  
diff-Pad\'e series analysis, 
rational number reconstruction,
elliptic functions, weight-1 modular forms.

\section{Introduction}
\label{pre}

Wu, McCoy, Tracy and Barouch~\cite{wu-mc-tr-ba-76} have shown 
that the magnetic susceptibility of the square lattice 
Ising model can be expressed
as an infinite sum of contributions,
known as {\em $n$-particle contributions}, so that the high-temperature 
susceptibility is given by
\begin{eqnarray}
\label{plus}
kT \cdot \chi_H(w) \, \, =  \, \, \, \sum \chi^{(2n+1)}(w) \,\, = \, \,\,
{{1} \over {s}} \cdot (1 - s^4)^{\frac{1}{4}} \cdot 
\sum \tilde{\chi}^{(2n+1)}(w)
\end{eqnarray}
and the low-temperature susceptibility is given by
\begin{eqnarray}
\label{chimoins}
kT \cdot \chi_L(w)\, \,  = \,\,\, \sum \chi^{(2n)}(w) \,\, = 
 \, \,\, (1 - 1/s^4)^{\frac{1}{4}} \cdot \sum \tilde{\chi}^{(2n)}(w)
\end{eqnarray}
in terms of the self-dual temperature variable $w=\frac{1}{2}s/(1+s^2)$,
with $s =\sinh(2J/kT)$. 

As is now well known~\cite{wu-mc-tr-ba-76}, the $n$-particle contributions
have an integral representation and are given by the $(n-1)$-dimensional 
integrals~\cite{pal-tra-81,yamada-84,nickel-99,nickel-00}
\begin{eqnarray}
\label{chi3tild}
\tilde{\chi}^{(n)}(w)\,\,=\,\,\,\, {\frac{1}{n!}}  \cdot 
\Bigl( \prod_{j=1}^{n-1}\int_0^{2\pi} {\frac{d\phi_j}{2\pi}} \Bigr)  
\Bigl( \prod_{j=1}^{n} y_j \Bigr)  \cdot   R^{(n)} \cdot
\,\, \Bigl( G^{(n)} \Bigr)^2, 
\end{eqnarray}
where\footnote[1]{The Fermionic term $\,G^{(n)}$ has several 
representations~\cite{nickel-00}.} 
\begin{eqnarray}
\label{Gn}
G^{(n)}\,=\,\, \prod_{1\; \le\; i\;<\;j\;\le \;n}h_{ij}, 
 \qquad 
h_{ij}\,=\,\,
{\frac{2\sin{((\phi_i-\phi_j)/2) \cdot \sqrt{x_i \, x_j}}}{1-x_ix_j}}, 
\end{eqnarray}
and
\begin{eqnarray}
\label{Rn}
R^{(n)} \, = \,\,\, \,  {\frac{1\,+\prod_{i=1}^{n}\, 
x_i}{1\,-\prod_{i=1}^{n}\, x_i}}, 
\end{eqnarray}
with
\begin{eqnarray}
\label{thex}
&&x_{i}\, =\,\,\,  \,  \frac{2w}{1-2w\cos (\phi _{i})\, 
+\sqrt{\left( 1-2w\cos (\phi_{i})\right)^{2}-4w^{2}}},  
\\
\label{they}
&&y_{i} \, = \, \, \,
\frac{2w}{\sqrt{\left(1\, -2 w\cos (\phi _{i})\right)^{2}\, -4w^{2}}}, 
\quad \quad  \quad  \quad \sum_{j=1}^n \phi_j=\, 0  
\end{eqnarray}
valid for small $w$ and, elsewhere, by analytic continuation. The variable $w$ 
corresponds to small values of $s$ {\em as well as} large values of $s$.
It is worth noting that the series expansions for $\, \tilde{\chi}^{(n)}$ 
in the variable $\, w$ have integer coefficients. From
the first $\, \tilde{\chi}^{(n)}$, the coefficients for generic $n$ can
be inferred~\cite{bo-bo-ha-ma-we-ze-09}
\begin{eqnarray}
\label{closedn}
&&\tilde{\chi}^{(n)} \, = \, \, 2^n \cdot w^{n^2} \cdot 
\Bigl(1 \, + \, 4 \, n^2 \cdot w^2 \,  
+ \, 2 \cdot (4\, n^4 \, +13\, n^2 \, +1)\cdot w^4 \, \, \nonumber \\
&& \qquad \quad \quad \quad  + \,
{{8} \over {3}} \cdot (n^2+4)\, (4\, n^4\, +23\, n^2+3) \cdot w^6
\, + \,  \cdots  \,  \Bigr),
\end{eqnarray}
where the coefficients are valid up to $\, w^2$ for $\, n\, \ge \, 3$,
 $\, w^4$  for $\, n\, \ge \, 5$ and  $\, w^6$
 for $\, n\, \ge \, 7$ (in particular
it should be noted that $\tilde{\chi}^{(n)}$ 
is an even function only for $n$ even).

In previous work~\cite{bo-gu-ha-je-ma-ni-ze-08} we performed massive computer
calculations to obtain the susceptibility of the square lattice
Ising model and the $\, n$-particle contributions $\, \tilde{\chi}^{(n)}$.
These calculations confirmed 
previously~\cite{bo-ha-ma-ze-07,bo-ha-ma-ze-07b}
conjectured singularities for
the linear ODEs of  $\, \tilde{\chi}^{(n)}$ (for $n=5$ and 6) 
and yielded the values of the associated
 local exponents. In addition some light
was shed~\cite{bo-gu-ha-je-ma-ni-ze-08}
 on important physical problems such as
 the existence of a {\em natural boundary} for 
the susceptibility of the square 
lattice Ising model and the subtle {\em resummation
of logarithmic contributions} from individual 
$\, \tilde{\chi}^{(n)}$'s resulting in the 
power-law behaviour of the full susceptibility $\chi$.

As far as the five-particle contribution to the susceptibility is concerned,
a long series $S(w)$ for $\, \tilde{\chi}^{(5)}$ 
was generated modulo the prime
$p_r=\, 2^{15}-19$ from which we  obtained~\cite{bo-gu-ha-je-ma-ni-ze-08} 
the corresponding Fuchsian differential equation. This Fuchsian linear 
ODE is of order 33 
and we denote by $L_{33}$ its linear differential operator. The 
calculation of the series is 
very time consuming and one cannot calculate (given presently available 
computational resources) the many series modulo various
primes required to reconstruct, through 
the {\em Chinese remainder} procedure, the exact
series for $\, \tilde{\chi}^{(5)}$, and, from this,
 deduce the corresponding exact Fuchsian
linear ODE. Our purpose here is, using 
the series and the linear ODE obtained modulo a single
prime, to perform, as far as possible, the factorization
of the linear differential operator 
$L_{33}$ and gain as deep an understanding as possible of the various 
factors occurring in its exact factorization (over the rationals).

In particular, we find that a certain linear
 combination of $\tilde{\chi}^{(1)}$ 
and $\tilde{\chi}^{(3)}$ can be removed 
from $\tilde{\chi}^{(5)}$ and the resulting 
series is a solution of an order 29 
linear ODE.  We develop methods which enable 
us to show that the corresponding linear 
differential operator $L_{29}$ splits into 
several factors and we present arguments that the
 order of any individual factor 
does not exceed five. The factor $L_5$ of
 maximum order  occurs as the left-most factor
of  $L_{29}$.  We show that $L_5$ is equivalent\footnote[9]{For the 
notion of equivalence of linear differential
operators see~\cite{Singer,vdP}.} 
 to the symmetric fourth power 
of~$L_E$, the linear differential operator corresponding to the complete 
elliptic integral~$E$, see~(\ref{eqKE}).
This result generalizes  what we have found 
in~\cite{ze-bo-ha-ma-04,ze-bo-ha-ma-05b,ze-bo-ha-ma-05}
 for the lower terms
$\tilde{\chi}^{(3)}$ and~$\tilde{\chi}^{(4)}$.  We
 therefore conjecture that a
linear differential operator $L_n$, 
equivalent to the symmetric $(n-1)$-th power 
of $L_E$, occurs as the left-most factor in the (minimal order) linear 
differential operators for all the $\tilde{\chi}^{(n)}$'s.

\section{Deciphering the structure of $\tilde{\chi}^{(n)}$: 
direct sums, symmetric powers and modular forms}
\label{deciphering}
A linear differential operator $L$  can be viewed formally as a 
non-commutative polynomial in $w$ and ${\rm D}_w$, 
where ${\rm D}_w\, = \, \,  \rmd/\rmd w$ 
is the derivation (or derivative) with respect to $w$.  
In previous works~\cite{ze-bo-ha-ma-04,ze-bo-ha-ma-05b,ze-bo-ha-ma-05} 
we have shown that the (minimal order) linear differential operators
for $\tilde{\chi}^{(3)}$ and $\tilde{\chi}^{(4)}$ (called respectively
$L_7$ and $L_{10}$) have a ``Russian-doll'' structure involving
the differential operators $L_1$ and $N_0$ for 
$\tilde{\chi}^{(1)}$ and $\tilde{\chi}^{(2)}$, respectively.
More precisely, $\tilde{\chi}^{(1)}$ and $\tilde{\chi}^{(2)}$ are solutions
of the linear ODEs corresponding to  $L_7$ and $L_{10}$, respectively.
In terms of linear differential operators this means that $L_1$ (resp. $N_0$)
right-divides $L_7$ (resp. $L_{10}$). Note that throughout this paper 
when we talk about a homogeneous linear differential equation and 
its associated differential operator we will use the terms ODE and 
differential operator, interchangeably.

One might then conjecture that this structure extends to the  linear differential
operator $L_{33}$ (for $\tilde{\chi}^{(5)}$) and the linear differential operator
$L_7$ (for $\tilde{\chi}^{(3)}$). We note that the singularities for the ODEs 
corresponding to~$\tilde{\chi}^{(3)}$ and $\tilde{\chi}^{(5)}$ are consistent with 
this assumption, that is {\em all} the singularities of $L_7$ also occur in $L_{33}$.
The check of the right division between operators  can be done simply by
generating the series $L_7 \left( S(w) \right)$ and obtaining the corresponding 
linear ODE. If the order of this latter linear ODE is less than 33, the assumption 
is verified, i.e. $L_7$ right-divides $L_{33}$. 
As reported in~\cite{bo-gu-ha-je-ma-ni-ze-08}
 this procedure leads to the factorization
\begin{eqnarray}
L_{33} \, = \,\,\,\, \, N_{26} \cdot L_7,
\end{eqnarray}
where $N_{26}$ is a linear differential operator of order 26.

A stronger conjecture amounts to saying that the linear differential
operator for $\tilde{\chi}^{(n)}$ occurs as part of a {\em direct-sum
decomposition} of the linear differential operator for $\tilde{\chi}^{(n+2)}$.
This conjecture is based on our
findings~\cite{ze-bo-ha-ma-04, ze-bo-ha-ma-05b, ze-bo-ha-ma-05} that the linear 
combinations $6\, \tilde{\chi}^{(n+2)} \, -n \, \tilde{\chi}^{(n)}$
 with $n=\, 1, \, 2$,  happen to verify a linear ODE
of lower order than the one for $ \tilde{\chi}^{(n+2)}$.
The conjecture was verified for $n=\, 3$ by
 obtaining~\cite{bo-gu-ha-je-ma-ni-ze-08} the
 minimal order Fuchsian linear ODE for
$6\, \tilde{\chi}^{(5)}\, -3\tilde{\chi}^{(3)}$, which
 happens to be of
order 30, so that
\begin{eqnarray}
\label{firstfactoL33}
L_{33} \, = \,\,\, \,\,   L_7 \, \oplus \, L_{30}.
\end{eqnarray}

This direct-sum structure, where a seventh order
linear differential operator combined with an order
30 linear differential operator gives rise to an order 33 
linear differential operator,
leads to the conclusion that there must be a fourth order
 linear differential
operator that right divides both $L_{30}$ and  $L_7$.
This kind of direct-sum structure (\ref{firstfactoL33}) is
 not seen in $L_7$ or $L_{10}$, the linear differential
 operators for $\tilde{\chi}^{(3)}$
and $\tilde{\chi}^{(4)}$.

Recalling~\cite{ze-bo-ha-ma-05c} the factorization of $L_7$,
the fourth order linear differential
 operator reads (following the notation 
of Eq.~(7) in~\cite{ze-bo-ha-ma-05c})
\begin{eqnarray}
B_{2} \cdot T_1 \cdot L_1 \, = \,\, B_{2} \cdot O_1 \cdot N_1 \,=\,\,
 X_1 \cdot Z_{2} \cdot N_1 \, =\, \, L_1 \, \oplus \, (Z_2 \cdot N_1)
\end{eqnarray}
and the factorization of $L_{30}$ in (\ref{firstfactoL33}) becomes
\begin{eqnarray}
\label{secondfactoL33}
L_{30} \, = \, \,\, 
 L_{26} \cdot (L_1 \, \oplus \, (Z_2 \cdot N_1)).
\end{eqnarray}
In a further step, Nickel\footnote[2]{We are grateful to B. Nickel for this result.}
has succeeded in showing that the differential operator $L_1$ (corresponding
to $\tilde{\chi}^{(1)}$) occurs via a direct sum in  $L_{30}$. This was done by 
considering the series for the combination 
\begin{eqnarray}
\label{nickelchi1}
\tilde{\chi}^{(5)} \,\,\, - {\frac{1}{2}} \,\tilde{\chi}^{(3)}\,\, \,
-\alpha \cdot \tilde{\chi}^{(1)}.
\end{eqnarray}
If a rational value of $\alpha$ can be found such that the resulting linear 
ODE has an order less than 30 (in fact equal to 29), then the direct sum 
assumption has been validated. This happens with $\alpha= \, -1/120$ 
and the final result is that the
combination
\begin{eqnarray}
\label{combinat}
\tilde{\chi}^{(5)}\,\, \, - {\frac{1}{2}}\, \tilde{\chi}^{(3)}\,\, \,
+{\frac{1}{120}} \, \tilde{\chi}^{(1)}
\end{eqnarray}
is annihilated by an order 29 linear differential operator
that we denote $L_{29}$, leading to conclude that 
\begin{eqnarray} 
\label{L30}
L_{30} \, = \, \, L_1 \oplus L_{29}.
\end{eqnarray}

At this point, guided by our results for the three terms $\tilde{\chi}^{(n)}$,
$n=\,\, 3,\, 4,\, 5$, one may wonder whether there is a common structure to 
the corresponding linear differential operators. 

Recall that the ODE for $\tilde{\chi}^{(3)}$ is of order seven and that
$\tilde{\chi}^{(3)}$ can be written as
\begin{eqnarray}
\tilde{\chi}^{(3)} \,=\, \,\,
 {1 \over 6} \, \tilde{\chi}^{(1)}\,\,  + \Phi^{(3)},
\end{eqnarray}
where $\Phi^{(3)}$ is a solution of a sixth order linear ODE.
 We thus have the {\em direct sum decomposition}
\begin{eqnarray}
L_7  \,=\,\,\,\, \,  L_1 \oplus L_6.
\end{eqnarray}
The sixth order operator $L_6$ has a third order linear 
differential operator $\,  Y_3$ as a left-most factor
\begin{eqnarray}
L_6  \,=\,\,\,\,  \, Y_3 \cdot Z_2 \cdot N_1,
\end{eqnarray}
and we have given the solutions of the linear ODE corresponding to $Y_3$
in~\cite{ze-bo-ha-ma-05c}. These solutions can be written~\cite{ze-bo-ha-ma-05c} 
as a homogeneous polynomial of the complete elliptic integrals $K$ and $E$ with 
homogeneous degree two (the order of $Y_3$ minus one).

Next we consider the tenth order linear ODE for $\tilde{\chi}^{(4)}$.
Recall that $\tilde{\chi}^{(4)}$ can be written as
\begin{eqnarray}
\tilde{\chi}^{(4)} \,=\, \,\,\, {1 \over 3} \tilde{\chi}^{(2)}\,\, \,
 + \,\Phi^{(4)},
\end{eqnarray}
where $\Phi^{(4)}$ is a solution of an eighth order linear ODE. 
We thus have the direct sum decomposition
\begin{eqnarray}
L_{10}  \,=\, \,\,  \,\, N_0 \oplus L_8.
\end{eqnarray}
The eighth order operator $L_8$ has a fourth order linear differential
operator $M_2$ as its left-most factor:
\begin{eqnarray}
L_8  \,=\, \, \,\, M_2 \cdot L_4.
\end{eqnarray}
The fourth order linear differential operator $L_4$ factorizes into four
first order linear differential operators as shown in Eq.~(42) 
of~\cite{ze-bo-ha-ma-05b}. As mentioned in~\cite{ze-bo-ha-ma-05c}  
(and now given explicitly in \ref{Sec:M2Sol}), the linear ODE corresponding to 
$M_2$ annihilates a homogeneous polynomial of $K$ and $E$ of (homogeneous) 
degree three, i.e. the order of $M_2$ minus one.

Similarly, we have shown that for $\tilde{\chi}^{(5)}$
\begin{eqnarray}
\tilde{\chi}^{(5)} \,=\,\, \,  \, {\frac{1}{2}} \,  \tilde{\chi}^{(3)}\,\, 
-{\frac{1}{120}}  \,  \tilde{\chi}^{(1)}\,\,  +\,\Phi^{(5)},
\end{eqnarray}
where $\Phi^{(5)}$ is annihilated by an order 29 linear ODE whose corresponding
differential operator we denote as $L_{29}$.

We conjecture that once the $\tilde{\chi}^{(n)}$ are depleted of the contributions 
$\tilde{\chi}^{(n-2k)}$  of lower index (with coefficients $\alpha_{n-2k}$, 
where  $\alpha_{n-2}=(n-2)/6$, and the remaining coefficients are to
be determined numerically)  
\begin{eqnarray}
\label{conject1}
\tilde{\chi}^{(n)} \,=\, \,\, 
\alpha_{n-2} \cdot  \tilde{\chi}^{(n-2)}\,\, \, 
+\alpha_{n-4} \cdot  \tilde{\chi}^{(n-4)}\, 
+\, \cdots\,\, \, +\,  \Phi^{(n)},
\end{eqnarray}
the  depleted series $\Phi^{(n)}$ will be solutions of linear ODEs of minimal order 
$q$, whose corresponding (minimal order) linear differential operators factorize as 
\begin{eqnarray}
\label{conject2}
L_q  \,\,=\,\,\,\, \,  L_n \cdot L_{q-n}, 
\end{eqnarray}
and where the linear ODE corresponding to the left-most factor $L_n$ 
has as a solution a homogeneous polynomial of complete elliptic integrals 
$E$ and $K$ of  degree $n-1$ (in other words $L_n$ is equivalent to the $(n-1)$-th 
symmetric power of $L_E$, annihilating $E$, see below).

This is what happens for the terms  $\tilde{\chi}^{(3)}$ and $\tilde{\chi}^{(4)}$.
One of the purposes of this paper is to show that this structure {\em also holds} 
for $\tilde{\chi}^{(5)}$. This amounts to demonstrating the occurrence of a 
fifth order linear differential operator $L_5$ at the left of $L_{29}$, with $L_5$  
being equivalent to the symmetric fourth power of the linear differential operator 
$L_E$ corresponding to the complete elliptic integrals $E$ (or $K$).

Before proceeding to show how we achieved this goal, some general 
remarks are in order. For an integral representation of a function
its series expansion $S(x)$ around the origin ($x=0$) is unique. 
This series can be annihilated by (is a solution of) many linear ODEs
of order $Q$ and degree $D$ (see Appendix B):
\begin{eqnarray}
        L_{QD} \,\,  = \,\, \, \,
 \sum_{i=0}^{Q} \Bigl(  \sum_{j=0}^{D}\,  a_{i j} \cdot  x^j \Bigr)
 \cdot \left( x\, {{\rmd} \over {\rmd x}} \right)^i. 
\end{eqnarray}

Among all these linear ODEs there is one of minimal order $q$  
and it is unique (its corresponding degree will be denoted by $D_0$). 
In terms of linear differential operators, 
the minimal order differential operator appears as a right-factor in the 
non-minimal order linear differential operators. The minimal order 
linear ODE may contain a very large number of apparent singularities 
and can thus only be determined from a very  large number of series 
coefficients (generally speaking $(q+1)(D_0+1)$ terms are needed).
Other (non-minimal order) linear ODEs, because they carry polynomials 
of smaller degrees, may require {\em fewer series coefficients} in order to 
be obtained. For any $Q > \, q$, a linear ODE annihilating  $S(x)$ 
(i.e. $L_{QD}(S(x))=\, 0$), can be 
found\footnote[1]{Of course the minimal order 
operator right-divides all these $L_{QD}$.} for $D$ sufficiently large, and if 
$Q$ is small enough we can choose $Q$ and $D$ such that 
$(Q+1)(D+1) \,< \, (q+1)(D_0+1)$. Among  the non-minimal linear ODEs 
there will generally be one requiring the
 {\em minimal number of terms}; in a 
computational sense one may view this as the ``optimal linear''
ODE\footnote[3]{The sizes (order and degree) of minimal order vs. optimal 
ODEs are very well understood in some particular cases.  For instance, the 
minimal order ODE (also called ``differential resolvent") satisfied by an 
{\em algebraic function} has coefficients whose degree is cubic in the degree 
of the function, while there exists a linear differential equation of order linear
in the degree whose coefficients are only of {\em quadratic} 
degree~\cite{bo-ch-le-sa-sc-03}. To our knowledge, analogous estimates do 
not exist yet for the (more general) case of linear ODEs satisfied by
 integrals of algebraic functions, such as $\tilde{\chi}^{(n)}$.}. 
In the case of $\tilde{\chi}^{(5)}$ this optimal linear ODE  can be discovered 
from some 7400 terms while the minimal order linear ODE requires almost
49100 terms. So when we consider, for instance, a linear differential operator 
such as $L_{29}$ (of minimal order 29), we are dealing in fact (as far as the
computations are concerned) with a much higher order linear differential operator.

The knowledge about the minimal order is ``inferred'' from many non-minimal 
order ODEs by using the remarkable formula~(\ref{ODEform}) below,
which we reported in~\cite{bo-gu-ha-je-ma-ni-ze-08}. Seeking a Fuchsian linear
ODE of order $Q$ and degree $D$ which annihilates a given series requires 
a certain number  of coefficients $N$. Formula~(\ref{ODEform})
{\em relates this number of required coefficients $N$ to the order $Q$ 
and degree} $D$ of the Fuchsian linear ODE. Remarkably, this ``ODE formula'' 
gives the number of required coefficients~$N$ as a linear combination of the
order $Q$ and degree $D$, while a naive and obvious upper bound for $N$
is  $(Q+1)(D+1)$. We denote by $\, f$ the difference between this naive 
upper bound and the  actual {\em number of required coefficients} $N$.

We have no proof of this formula, but it has been found to work 
\cite{bo-gu-ha-je-ma-ni-ze-08} for all the cases we have considered
\begin{eqnarray}
\label{ODEform}
N \, = \, \, \, d \cdot Q \, \,  + q \cdot D \, \,  -C \,=  
\,  \, \, (Q+1)(D+1)\, \,  -f.
\end{eqnarray}
This ODE formula is revisited in greater detail in \ref{Sec:ODEformula}
where all its parameters have now been understood. In all cases we have 
investigated, the parameter $q$ appearing in (\ref{ODEform}) is the 
minimal order of the linear ODE that annihilates $S(x)$.
The parameter $d$ is the {\em number of singularities}
 (counted with multiplicity) 
excluding any apparent singularities and the ``true'' singular point $x=\, 0$, 
which is already taken care of  by the use of the (so-called homogeneity) 
 differential operator $x {\rmd \over \rmd x}$. Finally we note that the degree 
of the apparent polynomial of the minimal order linear ODE (as well as the 
other parameters $d,\, q,\, C$ and~$f$) can be extracted exactly without 
obtaining the minimal order linear ODE (see (\ref{Dapp}) in \ref{Sec:ODEformula}).
 
As stated above we are dealing with linear differential
operators of higher orders than the minimal order and whose coefficients
are known modulo a prime. To factorize such large order linear differential
operators, we make use of the method sketched in Section~\ref{Sec:FacExp}.
This is done by ``following'' the series pertinent to a specific local exponent
at a given singular point. Linear combinations of series with different local
exponents can be studied as well. Our approach is similar to the
 one proposed by~van Hoeij in~\cite{vanHoeij-97} (a ``local'' factorization 
deduced from formal series analysis around each singularity followed by a
``Hermite-Pad\'e approximation'' to 
obtain the ``global'' factorization).
The main difference is\footnote[9]{Note that the DFactor routine 
from the DEtools Maple package, corresponding to an implementation of 
these {\em local-to-global} ideas~\cite{vanHoeij-97}, fails to factor
$\, L_6$ of $ \, \tilde{\chi}^{(3)}$. The method 
we display in Section~\ref{Sec:FacExp} 
actually succeeds in finding this factorization.} that in our case the operators
to be factorized are defined over a field of 
positive characteristic\footnote[5]{Note 
that in principle one could also resort to algorithms specially dedicated to 
factoring linear differential operators modulo a prime~$p$~\cite{cluzeau-03}. 
However, at present these algorithms are far from being efficient enough to handle 
factorizations modulo primes as big as $p_r = \, 2^{15}-19$.}.

Actually, the modular nature of our calculation is of great help in this since, with 
the coefficients being known modulo a prime, 
the coefficients in the linear combination
of solutions with given local exponents can 
take only a finite number of integer values,
so that ``guessing" the correct combination can be  done by exhaustive search. For 
each series used as a candidate to ``break"  the linear differential operator  under 
consideration we compute three (or more)  linear ODEs and from the ODE 
formula~(\ref{ODEform}) {\em we infer the minimal order}.

Another point that we address in this paper is the ``complexity" of the linear 
differential operators corresponding to the $\tilde{\chi}^{(n)}$ seen through the 
various factors occurring in the factorization. One notices that for $\tilde{\chi}^{(4)}$, 
the factors are either of order one, or are symmetric powers of the linear differential 
operator $L_E$. In contrast, the linear  differential operator for $\tilde{\chi}^{(3)}$ 
contains a factor $Z_2$ of order two which is not equivalent to the linear differential 
operator $L_E$. Recently it has been shown~\cite{bo-bo-ha-ma-we-ze-09} 
that the solution of the linear ODE corresponding to $Z_2$ is a hypergeometric 
function (up to a Hauptmodul pull-back) corresponding to a {\em weight-1 modular form} 
(see~\cite{maier-05}). We think it is important to examine whether  the linear ODE 
for $\tilde{\chi}^{(5)}$ contains other factors, besides various factors equivalent to 
symmetric powers of $L_E$,  such as the factor $Z_2$ occurring for 
$\tilde{\chi}^{(3)}$, that may have a {\em modular form interpretation}.

Finally we wish to emphasise that the linear differential operator $L_{33}$ is 
globally nilpotent\footnote[3]{The mathematical
concept of global nilpotence while quite formal is nevertheless easy to explain. 
Firstly, if $p$ is a fixed prime number, then 
the differential operator $L$  is said to have nilpotent $p$-curvature 
iff modulo $p$, it right-divides the pure
 power ${\rm D}_w^{p\cdot{\rm ord}(L)}$ of the 
derivation (${\rm D}_w\, = \, {\rm d}/{\rm dw}$). Secondly, $L$ is called
 ``globally nilpotent" if it has nilpotent $p$-curvature 
modulo  almost all prime numbers $p$ (all primes except
 a finite number of prime).} since it corresponds
 to a linear ODE that annihilates an integral 
of an algebraic integrand (\ref{chi3tild})
  (it is {\em ``derived from geometry''},  
DFG, see~\cite{bo-bo-ha-ma-we-ze-09} and references
 therein). While this paper is not concerned with global 
nilpotence as such it must be emphasized that the nilpotent condition places
very severe restrictions on a linear differential operator, and in particular,
{\em provides a proper framework}\footnote[3]{The Cauchy-Peano 
theorem~\cite{Poole},  which guarantees 
the existence of series-solutions
in the classical study of ODEs,
 {\em does not apply to linear differential 
equations in positive characteristic}!
 This means that in general,  a linear differential 
equation considered modulo a prime number $\, p$
 {\em does not admit a basis of power series
 solutions modulo $\, p$, even at an ordinary point}.
} {\em for the existence of basis of series solutions modulo primes}
 for the ODEs (see Theorem 4 in~\cite{honda-81}). Furthermore,
  we use two important
consequences of the global nilpotence of $L_{33}$. Firstly, globally nilpotent
operators are necessarily Fuchsian and permit only rational local exponents.
Secondly, if $L$ is globally nilpotent so is any factor of $L$.
 
\section{Working with non-minimal order linear differential operators}
\label{Sec:NonMin}
Once one introduces the particular linear combination (\ref{combinat}) of
$\, \tilde{\chi}^{(1)}$ and $\, \tilde{\chi}^{(3)}$ and (modulo the prime $p_r$) of 
$\, \tilde{\chi}^{(5)}$, it is sufficient
 to focus on the resulting series and its linear 
ODE (with the operator $L_{29}$). From the  linear ODE for $\, \tilde{\chi}^{(5)}$ 
it is straightforward to obtain the (linear) recursion for the series coefficients and 
using the combination (\ref{combinat}) calculate as many terms\footnote[1]{Note 
that the number of coefficients of all the series used in our calculations does not 
exceed the value of the prime $p_r$.} as required for $\Phi^{(5)}$. It is thus not 
difficult to obtain minimal order ODEs requiring fewer terms than $p_r$.
We continue however, as in~\cite{bo-gu-ha-je-ma-ni-ze-08}, to work
with non-minimal order ODEs for which fewer series terms are needed than for 
the minimal order ODE. In particular, we make use of the ODE formula 
(see \ref{Sec:ODEformula}) to infer the order and degree of the minimal order 
ODE. This formula also enables us to {\em control the minimum number of series 
terms} necessary to find a Fuchsian linear 
differential  equation (of a given order $Q$
and degree $D$) which annihilates the series. In the sequel, when we say that a
linear ODE of order~$q$  has been obtained, we mean that we have obtained 
at least three non-minimal order ODEs and that the minimal order $q$ has been 
inferred from the ODE formula. 

From the series (\ref{combinat}) one can obtain many non-minimal
order  linear ODEs and the resulting ODE formula for $L_{29}$ reads
\begin{eqnarray}
\label{magic}
N \, = \, \, \, 68\, Q \,\,   + 29\, D  \,\, -706 \,\,\,
=\, \,\,\, \, (Q+1)(D+1) \,\, -f.
\end{eqnarray}
Our understanding of the ODE formula (see \ref{Sec:ODEformula}
 and in particular (\ref{Dapp})) enables us to find $D_{app}=\, 1169$ 
as the degree of its apparent polynomial for the minimal order operator $L_{29}$ 
without actually producing this  minimal order operator.

In~\cite{bo-gu-ha-je-ma-ni-ze-08} we found that there is a
simple rational solution of the linear ODE corresponding to $L_{30}$
(and now also $L_{29}$) which is the square of $\tilde{\chi}^{(1)}$,
\begin{eqnarray}
\label{chi1square}
\left( \tilde{\chi}^{(1)} \right)^2 \, = \,\,\, \,\,
 {\frac{w^2}{(1-4w)^2}}, 
\end{eqnarray}
whose corresponding linear differential operator we denote $L_1^s$.

In this paper we use linear ODEs that are {\em not} of minimal order 
to represent the minimal order linear ODE annihilating a given series.
We also compute the local exponents at various singular points of the
non-minimal linear ODE and consider them as local exponents of the 
minimal order linear  ODE. A remark is in order here. The local exponents 
at $w=\, 0$ of the linear ODE corresponding to $L_{29}$ 
are\footnote[2]{Throughout the paper the multiplicity of an exponent is 
denoted by a power: $2, \,2, \,2, \, 2$ $\, \, \rightarrow \, \, 2^4$.}
\begin{eqnarray}
\rho \, =\,\,\,\,\,
 1^5,\,\,\, 2^4,\,\,\, 3^4,\,\,\, 4^3,\,\,\, 5^3,\,\, \,6^3,\,\,\, 7^2,\,\,\, 
8,\, \,\,9^2,\,\,\, 10,\,\,\, 12.
\end{eqnarray}
\noindent
In our computation, the non-minimal linear ODE that represents $L_{29}$ is of 
order~51. One should then really obtain 51 local exponents. It so happens that 
the 22 ``spurious" exponents appear in the indicial equation as roots of polynomials 
in $\rho$ of degree two and higher. These exponents are {\em not rational} 
(indicial polynomials modulo a prime of degree higher than one and irreducible) 
and therefore cannot be local exponents for $L_{29}$, which is  globally nilpotent and, 
hence, has only rational exponents~\cite{bo-bo-ha-ma-we-ze-09}.
Had the indicial equation of the non-minimal linear ODE given more than 29 
rational exponents then we would have had to produce other non-minimal 
linear ODEs and obtain the local exponents of $L_{29}$ as those common to 
all the non-minimal ODEs.

\section{Factorization of differential operators
 versus local exponents}
\label{Sec:FacExp}

Let us start by giving a brief overview of our 
factorization procedure. For a linear ODE 
of order $q$ (which may be known exactly or modulo a prime) we compute the local
exponents at  a given singular point $w=\, w_s$
 (such as the origin). We create a series 
$S_p(w)$ starting with the highest integer exponent $n_p$ (we seek
mainly to utilise only those solutions analytic 
at the origin). This series can be obtained 
to arbitrary length (though shorter than the prime in use) in linear time since 
we have the linear ODE and, hence, the recursion for the series coefficients.
We then check to see whether the particular solution $S_p(w)$  is the solution of
a linear ODE of order less than $q$. If so, the procedure is repeated for
each new factor in turn. If not, we generate the series $S_{p-1}(w)$ 
starting at the second highest exponent $n_{p-1}$. The series $S_{p-1}(w)$ contains, 
via a free parameter, a linear  combination of the solution $S_p(w)$.
We then let the free coefficient of the linear combination 
vary over the whole (finite) interval $[1,p_r]$, 
given by the prime $p_r$ we are using, 
until a linear ODE of order less than $q$ (if
 such an ODE exists) is found. And then 
the procedure is repeated.

For a linear ODE of order $q$, let $L_q$ denote the corresponding differential 
operator. Consider a singular point $w=\, w_s$ (for instance $w_s=\, 0$) and
assume the local exponents at this point are
\begin{eqnarray}
\rho_1^{m_1}, \, \rho_2^{m_2},
 \cdots, \rho_p^{m_p}, \qquad \qquad \sum_{j=1}^p m_j \, =\,\,   q, 
\end{eqnarray}
where  $m_j$  is the multiplicity of the exponent $\rho_j$.
In our cases the exponents are either integers or rational numbers. Here 
we  utilize only solutions, which are analytic at the singular point $w_s$.
So in what follows we consider only integer exponents and we denote
these as $n_p$.  We can then plug the series 
\begin{eqnarray}
S_p (w)\,=\,\, \,  w^{n_p}\,\,   + \sum_{k\geq n_p+1} a_k\, w^k, 
\end{eqnarray}
into the linear ODE. Demanding $L_q \left( S_p(w) \right)=\, 0$ will fix all  
coefficients $a_k$. By producing enough terms we can find the linear ODE 
for the particular series solution $S_p(w)$, which is by construction a 
solution of $L_q$. The resulting ODE will
 either have order $q$ or order $q_1 <\, q$.
In the first case this could mean that $L_q$ is irreducible, or
$L_q$  does factorize but the factor ``responsible" for 
annihilating the solution $S_p(w)$ is a left-most factor.
In the second case we have the factorization
\begin{eqnarray}
\label{LqLqq}
L_q \,\,=\,\,\,\,  L_{q-q_1} \cdot L_{q_1}. 
\end{eqnarray}
To summarise, the series corresponding to the highest local exponent
leads to either the full ODE or to a ``breaking" of the original ODE. 

If the series $S_p(w)$ (corresponding to the highest local
exponent) reproduces the full linear ODE we turn to the second
highest exponent  $n_{p-1}$.  In this case, a series starting as
$w^{n_{p-1}}\,+\cdots$, plugged into the original linear 
ODE, will yield the expansion
\begin{eqnarray}
\label{Sp-1}
S_{p-1}(w) \,=\,\,\,  w^{n_{p-1}}\,
 +  \sum_{k\geq n_{p-1}+1}^{n_p-1} a_k\, w^k 
+ a_{n_p}\, w^{n_p}\,  + \sum_{k\geq n_p+1} c_k\, w^k
\end{eqnarray}
where all $a_k$ up to (but not including) $a_{n_p}$ are fixed and the $c_k$'s 
depend linearly on the free coefficient $a_{n_p}$, i.e. $S_{p-1}$(w) is a 
one-parameter solution. The series $S_{p-1}(w)$ is a sum of a series
starting as $w^{n_{p-1}}+\cdots$ and the series\footnote[3]{Alternatively
we can view  this procedure as looking at a linear combination of the two formal 
series solutions starting as
 $w^{n_{p-1}}\, + \cdots\, $ and $w^{n_{p}}\,+\cdots\, $
respectively.}  $a_{n_p}\,S_p(w)$.   For generic values of the coefficient
$a_{n_p}$ the series $S_{p-1}(w)$ will give rise to the full linear ODE.
But for some values of the coefficients $a_{n_p}$, the series $S_{p-1}(w)$
may be the solution of a linear ODE 
of order less than $q$. This is what leads to the
factorization of $L_q$. Intuitively we may hope that such a procedure
can work for the following reason. If the original operator has many smaller
factors this would indicate that there is a basis of solutions much simpler
than those requiring the full ODE. We don't know this basis but by taking
a linear combination of two formal ``full'' solutions (which obviously are linear 
combinations of the basis solutions) it is possible that we can find
values of the combination coefficients such that the resulting series is a
solution of a simpler ODE (for these special values some of the basis 
solutions from the two formal solutions cancel each other).

Similarly, the series solution of $L_q$ that starts
 at the third highest exponent
$n_{p-2}$ will be a two parameters solution (for simplicity we assume
that the exponents differ by one)
\begin{eqnarray}
S_{p-2}(w) \,=\,\,\,  w^{n_{p-2}}\, + a_{n_{p-1}}\, w^{n_{p-1}}\,
 + a_{n_p}\, w^{n_p}\, 
 + \sum_{k\geq n_p+1} c_k\, w^k
\end{eqnarray}
where the $c_k$ depend linearly on both the free coefficients $a_{n_{p-1}}$
and $a_{n_p}$.

To demonstrate how the procedure works in practice
we consider the seventh order linear ODE for $\tilde{\chi}^{(3)}$
\cite{ze-bo-ha-ma-04,ze-bo-ha-ma-05} (denoted $L_7$) .
At the singular point $w=0$, the local exponents are
\begin{eqnarray}
  \,\, 1^3,\,\,  2^2,\, \, 3,\, \, 9.
\end{eqnarray}
Acting with the linear ODE for $\tilde{\chi}^{(3)}$ on the series
that starts as $w^9\, + \cdots$, (i.e. with the highest exponent)
\begin{eqnarray}
S_9(w) \,=\,\,\,   w^{9}\, \,  + \sum_{k\geq 10} a_k\, w^k, 
\end{eqnarray}
fixes all the coefficients. We thus obtain the expansion
at $w=\, 0$ of $\tilde{\chi}^{(3)}/8$, leading to the full linear ODE.
Of course, this is not surprising, since the series for $\tilde{\chi}^{(3)}$
used to ``generate" the linear differential operator $L_7$ starts as $w^9$
as per (\ref{closedn}), so the unique series $S_9(w)$ must be proportional to 
 $\tilde{\chi}^{(3)}$. The series $S_9(w)$ cannot be used to
 ``break'' $L_9$,  since this is the minimal order operator annihilating
$\tilde{\chi}^{(3)}$. 
 
Consider next a series that starts as
 $w^3 \, + \cdots$, i.e. with the second highest exponent
\begin{eqnarray}
S(w) \,=\,\,  w^{3}\,  + \sum_{k\geq 4} a_k\, w^k.
\end{eqnarray}
We insert this series into the exact ODE for $\tilde{\chi}^{(3)}$  and
 then we solve (term by term) the equations arising from
 $L_7 \left( S_3(w) \right) =0$. Doing this we find that the coefficients 
 $a_4$, $a_5$, $\cdots$, $a_8$ have to be fixed while the coefficient $a_9$ 
 remains undetermined and hence enters the series as a free parameter. 
 The remaining coefficients are all given in terms of $a_9$.  
 \begin{eqnarray}
&&S(w) \,= \, \, w^{3} + 3w^4 +22w^5\, +74w^6\, +417w^7\, 
+1465w^8\, + a_9 w^9  \\
&&\quad \,   +26839w^{10}+(36a_9-139067) \cdot w^{11}
+(4a_9 +443325) \cdot w^{12}\, + \, \cdots \nonumber 
\end{eqnarray}
The terms in $S(w)$ in front of the free coefficient $a_9$ 
are  the coefficients of the series $S_9(w)$. We define $S_3(w)$
to be the series obtained from $S(w)$ by setting $a_9=\, 0$.  
In order to break the operator $L_7$ we look at linear combinations 
$S_{\alpha}(w)=\, S_3(w)\, +\alpha S_9(w)$. For generic values of 
$\alpha$ the series $S_{\alpha}(w)$ is annihilated only by the full
 ODE of order seven. However, it is possible that for special
values of $\alpha$ the series $S_{\alpha}(w)$ is the solution of
a linear ODE of order less than seven. 

We do not know if the ``splitting" values of $\alpha$ can be obtained except 
by a ``guessing'' procedure. The use of modular calculations  is very useful 
in the search for the special splitting values. The series $S_9(w)$ and $S_3(w)$
can be obtained modulo any prime $p_r$ and in the modular calculations $\alpha$ 
can take its value only in the finite range
 $[1, p_r]$. If a rational splitting value of 
$\alpha$ exists it can be found by looking for
 an underlying ODE of order less than 7
annihilating the series $S_{\alpha}(w)$. In the search we use the optimal ODE,
which is of order 10 and degree 19 with $N=213$. 
We used the prime $p_r=\, 32749$ in our search. For
 each value of $\alpha \in  [1, 32749]$
we calculated the series modulo $p_r$ and then looked for an annihilating ODE 
of order 10 and degree 19. For any value of $\alpha$ such an ODE exists and
for almost all values $N=213$. However, for the
 special values $\alpha =7463$  and 7467
we have $N=140$  and 206, respectively. The decrease in $N$ is a sure sign that
a simpler ODE annihilates $S_{\alpha}(w)$. In this particular case we find that
the ODE for $\alpha = \, 7463$  is of order four while for  $\alpha =\, 7467$ the
ODE is of order six. 

 In the  case $\alpha = \, 7463$  the linear differential operator is 
 $X_1 \cdot Z_2 \cdot N_1 =\, B_2 \cdot O_1 \cdot N_1 =\, B_2 \cdot T_1 \cdot L_1$, 
 while in the case  $\alpha =7467$  the linear differential operator 
 is  $Y_3 \cdot Z_2 \cdot N_1$.  These linear differential operators are 
 factors of $L_7$ that were already found in~\cite{ze-bo-ha-ma-05c} (the indices
 indicate the orders of the corresponding linear differential operators)
\begin{eqnarray}
\label{factoL7}
&&L_7 \,= \, \, M_1 \cdot Y_3 \cdot Z_2 \cdot N_1 \, =\, 
B_3 \cdot X_1 \cdot Z_2 \cdot N_1 \nonumber \\
&& \quad \quad = \, B_3 \cdot B_2 \cdot O_1 \cdot N_1 \, = \, 
B_3 \cdot B_2 \cdot T_1 \cdot L_1
\end{eqnarray}

\vskip .1cm 
{\bf Remark}: We note that the method is not special to
formal series. \textit{A fortiori}, it applies to linear combinations of solutions
{\it suspected} of being parts of a direct sum.
For instance, removing the series
$\alpha \, \tilde{ \chi^{(1)}}$ from
the series (modulo a prime) $\tilde{\chi}^{(5)}-{1 \over 2} \tilde{\chi}^{(3)}$ 
(see (\ref{nickelchi1})),  and, letting $\alpha$ vary in the interval
$[1, p_r]$ (recall that $\alpha$ which is a rational number appears as an
integer modulo a prime), will give (for one value of $\alpha$)  a linear ODE of
order~29 if the linear differential operator $L_1$ for $\tilde{\chi}^{(1)}$
is in a direct sum in $L_{30}$,
the linear differential operator for 
$\tilde{\chi}^{(5)}\, -{1 \over 2} \tilde{\chi}^{(3)}$ .
If $L_1$ had not been part of a direct sum the outcome would have been an order
30 linear ODE for {\em all} values of  $\alpha$.

\section{Factorization modulo a prime of the linear differential operator $L_{29}$}
\label{Sec:FactL29}

We turn now to the factorization of $L_{29}$
for which we know that $L_1^s \oplus (Z_2 \cdot N_1)$ is a factor.
We focus solely on the analytical solutions at $w=\, 0$ and we first produce
the unique series that starts as $S_{12}(w)=\, w^{12}\,  + \,\cdots$,  where the
coefficients in $S_{12}(w)$ are given by $L_{29}(S_{12}(w)) =\, 0$.
We found that $S_{12}(w)$ is the solution of an order nine linear ODE 
(with linear differential operator $L_9$) with ODE formula:
\begin{eqnarray}
N \, = \, \, \, 18\, Q\,\, + 9\, D\, \,-73 
\,\,\,=\,\,\,\,\, (Q+1)(D+1)\,\,-f.
\end{eqnarray}
\noindent
We know that both $\left( \tilde{\chi}^{(1)} \right)^2$ 
and the solutions of the 
linear ODE corresponding to $Z_2 \cdot N_1$ occurring in $L_7$, should be 
in $L_{29}$. By explicit checking we found that  only $L_1^s$ and $N_1$ are 
factors of $L_9$.  We can then add the solutions 
of $Z_2 \cdot N_1$ to the solutions 
of $L_9$  to produce an 11th order linear ODE (denoted $L_{11}$). At the operator 
level this is done by a direct-sum construction 
 $L_{11}=\,  L_9 \oplus (Z_2 \cdot N_1)$. 
At the series level used in the ODE search programs, it can be  done by creating a 
``generic" solution of $L_9$  (generic means a series that  gives the full ODE) and 
then forming a linear combination with a generic solution of $Z_2\cdot N_1$  to
produce a series which is a generic solution of  $L_{11}$.
The resulting linear differential operator $L_{11}$ has the  ODE formula: 
\begin{eqnarray}
N \, = \, \,\, \, 24\, Q \,\, + 11\,\, D \,\, -111
\,\, \,= \,\,\,\, \,(Q+1)(D+1)\,\,-f.
\end{eqnarray}
\noindent
We have thus shown the following factorization of $L_{29}$
\begin{eqnarray}
\label{firstfactoL29}
L_{29}\, \,=\,\,\,\,\,\, L_{18} \cdot L_{11}.
\end{eqnarray}


Before proceeding  we wish to clarify the meaning of the ODE
corresponding to the left factors. Having obtained the linear differential 
operators $L_{29}$ and $L_{11}$, a right division should give the linear 
differential operator $L_{18}$. One should bear in mind that the order of 
these operators is large and our representation of them are not of minimal order.
In the computation, the linear differential operators representing $L_{29}$, 
$L_{18}$ and $L_{11}$ are of orders 51, 32 and 17, respectively.
Our representation of the factorization  (\ref{firstfactoL29}) reads in fact
\begin{eqnarray}
O_{22} \cdot L_{29} \, \,\,=\,\,\,\, \, (O_{14} \cdot \tilde{L}_{18})
\cdot (O_6 \cdot L_{11})
\end{eqnarray}
where the equality stands for ``both sides acting on $S(w)$ give zero".
Since the series solution $S(w)$ {\em demands} an order 29 linear ODE, and 
the order of the extra operator $O_6$ is arbitrary, 
there are intertwinners leading to
\begin{eqnarray}
O_{22} \cdot L_{29} \,=\,\,\,\, O_{14} \cdot \tilde{O}_6 \cdot
\left( L_{18} \cdot L_{11} \right)
\end{eqnarray}
With the relation 
$\tilde{L}_{18} \cdot O_6\,  =\,\,   \tilde{O}_6 \cdot L_{18}$,
the linear differential operators $\tilde{L}_{18}$ and $L_{18}$ are
equivalent
and have the same factorization structure.
 

Next we take the series
 $S(w)=\, \tilde{\chi}^{(5)}\,\, - \tilde{\chi}^{(3)}/2 \,\,
 + \tilde{\chi}^{(1)}/120$ and plug it
 into the linear ODE for $L_{11}$ to produce
a new series whose linear ODE corresponds to the linear differential
operator   $L_{18}$.
This linear ODE has the formula:
\begin{eqnarray}
N \, = \, \, \, 44\, Q\,\, + 18\, D \,\,+873\, \,=\,
 \,\,\,\,  (Q+1)(D+1)\,\,-f. 
\end{eqnarray}
\noindent
To proceed further with the factorization, we 
compute the local exponents
at $w=\, 0$ for the linear ODE corresponding to $L_{18}$. These are:
\begin{eqnarray}
\rho \, =\,\,\,\,\, 
 1^3,\,\,\, 2,\,\,\, 3^2,\,\,\, 4,\,\,\, 5^2,\,\,\, 
6^3,\,\, \,7^2,\,\,\, 8,\,\,\, 9^2,\,\,\, 10.
\end{eqnarray}
For the linear differential 
operator $L_{18}$ we look at the solution that 
starts as $S_{10}(w)=\, w^{10}\, + \cdots$ 
Unfortunately, this gives linear ODEs
 with the same ODE formula as $L_{18}$, that is
the series reproduces the complete linear ODE represented by $L_{18}$.
This means that the factor  responsible for the 
annihilation of $S_{10}(w)$
occurs at the left of $L_{18}$.

The second highest exponent is $\rho=9$. When
 $L_{18}$ is applied to a generic series 
$S_{9}(w)=\, w^{9}\, + \, \cdots$ we obtain a series that depends
on the coefficient in front of $w^{10}$. This 
one-parameter series starts
as (modulo the prime $p_r$)
\begin{eqnarray}
w^9\, + a_{10}\, w^{10}\, 
+\, (15419\,a_{10}\,+10040) \cdot  w^{11}
\, + \,  \cdots
\end{eqnarray}
The series, collected in $a_{10}$, enables us to reconstruct 
 the full linear differential operator $L_{18}$, but it is
possible that the above combination may 
give a linear operator of {\em smaller order}
for particular values of $a_{10}$, which 
have to be found by experimentation.
It is here that the modular calculations
 allow us to find a definitive answer. 
Modulo the prime $p_r$, the coefficient $a_{10}$
spans a finite set of integer values $[1, p_r ]$.
The determination of the coefficient $a_{10}$
 is thus feasible in a
finite computational time by exhaustive search.

With the value $a_{10}=\,12999$ we found that 
 the series $S_9$ gives
a linear ODE of {\em smaller order}
 than $L_{18}$ with ODE formula
\begin{eqnarray}
N \, = \, \, \, 
36\, Q\,\, + 13\, D\,\, +715 
\, \, \,=\,\,\,\, (Q+1)(D+1)\,\, -f.
\end{eqnarray}
The local exponents at $w=\,0$ for this  linear 
ODE of order 13 (that we denote $L_{13}$) are:
\begin{eqnarray}
\rho \, =\,\,\,  \, \,
1^2,\,\, \,\, 2,\,\, \,\, 3,\,\, \,\, 4,\,\, \, \,5^2,
\,\, \,\, 6^2,\,\,\, \, 7^2,\,\,\, \, 8,\,\,\, \, 9.
\end{eqnarray}
The highest exponent is indeed nine, of which the
associated solution gave us $L_{13}$.

Let us be more explicit on the meaning of the combination
$w^9 \, + 12999\,w^{10}\, +\, \cdots$ that has 
given rise to the 13th order linear ODE.
By ``following'' the series $w^{10}\, + \cdots$ 
we obtained the full linear ODE.
The factor that annihilates this series is 
thus to the left in the factorization of $L_{18}$. We find that this
series comes with a $\log(w)^4$ behaviour (see below).
By the combination $w^9 \, + 12999\,w^{10} \, +\cdots$,
 we are looking for the particular value of $a_{10}$ that 
{\it removes} this logarithmic solution from the linear ODE corresponding 
to $L_{18}$. Thus the linear ODE for $L_{13}$ no
longer has a solution behaving as $\, \, w^{10} \, \log(w)^4$.

Completing from $L_{13}$ to $L_{18}$ we obtain a fifth order linear ODE
(called $L_5$) with ODE formula:
\begin{eqnarray}
N \, = \, \, \, 
8\, Q \,\,+ 5\, D \,\,+912 \,  \,\,=\,\,\,  \,(Q+1)(D+1)\,\,-f.
\end{eqnarray}
We thus have the factorization:
\begin{eqnarray}
L_{29}  \,\,=\,\,\,\, \, L_5 \cdot L_{13} \cdot L_{11}.
\end{eqnarray}
The fifth order linear differential operator $L_5$ 
is the one whose existence we conjectured previously and 
which we believe should annihilate a homogeneous polynomial of 
the complete elliptic integrals $\, E$ and $\, K$
of (homogeneous) degree four.

The local exponents at the origin of the linear 
ODE corresponding to $L_5$ are
\begin{eqnarray}
w =\,  0, \quad \quad \quad  \quad 
\rho\,=\, \,\,\,1,\,\, \,3,\, \,\,6,\,\,\, 9,\,\,\, 10.  
\end{eqnarray}
Plugging a generic series $\sum c_n\, w^n$  into the linear ODE fixes all the
coefficients (including $c_1$, $c_3$, $c_6$, $c_9$) with the exception of the
coefficient $c_{10}$.  The ``survival" of
  a single coefficient is a particular feature 
of an irreducible factor with one non-logarithmic solution.
The exponents at the other singularities 
(apart from $w= \, \infty$) are: 
\begin{eqnarray}
&w \,  = \,  \,  1/4, \quad \quad \quad  \quad 
&\rho =\,\, \,-29^2,\,\,\, -28,\,\,\, -23,\,\, \,0, \nonumber \\
&w  \, = \,  \,  -1/4, \quad \quad \quad  \quad 
& \rho = \,\,\,-35^2,\,\,\, -33,\,\,\, -31,\,\,\, 0. \nonumber 
\end{eqnarray}
This suggests that one should plug  the following ansatz into the linear ODE:
\begin{eqnarray}    
\label{eqKE}
{\frac{1}{(1-4w)^{29} (1+4w)^{35}}} \cdot 
 \sum_{i=0}^4 \, P_{4-i, i} \cdot  K^{4-i} \, E^i
\end{eqnarray}
where $K$ and $E$ denote the complete elliptic integrals
\begin{eqnarray}
K  \, =\,\, _2 F_1 ([1/2, 1/2], [1], 16w^2), \quad \quad
E  \, =\,\, _2 F_1 ([1/2, -1/2], [1], 16w^2). \nonumber
\end{eqnarray}
Collecting terms of the form $K^{4-i} \, E^i$ we determine the polynomials  
$P_{4-i, i}$ whose degrees are increased steadily
until we obtain a solution\footnote[5]{Once the solution has
been obtained a check to any order can be 
carried out. Typically a good check amounts 
to plugging polynomials of degree 300 into (\ref{eqKE}).}.
With degree  around 200 the following solution was found
\begin{eqnarray}
\fl \qquad \qquad {\frac{w}{(1-4w)^{29} (1+4w)^{35}}} \cdot 
\Bigl((1-16w^2)^3\,  P_{4,0} \cdot K^{4} \,\, 
+(1-16w^2)^2 \, P_{3,1} \cdot  K^3\,E 
\nonumber \\
 \qquad \qquad + (1-16w^2)\,  P_{2,2} \cdot  K^2 \, E^2 \,
 + P_{1,3} \cdot K \,E^3 \, + P_{0,4} \cdot E^4 \Bigr),
 \nonumber
\end{eqnarray}
where $P_{4-i, i}$ are polynomials in $w$ 
with coefficients known modulo the
prime $p_r$ and of degree respectively 200, 202, 204, 204 and 204.
The expressions for the polynomials $P_{4-i, i}$  can be found
in~\cite{WebJensen}. As conjectured
the linear differential operator $L_5$ {\em is thus equivalent to 
the symmetric fourth power of} $L_E$.

\subsection{The linear differential operator $L_{11}$ has six factors}
\label{Sec:L11}
As shown above the linear differential operator $L_{29}$ factorizes into 
three factors of order 11, 13 and 5. We have just shown 
that the fifth order linear differential operator is irreducible. Next we consider 
the linear differential operator $L_{11}$.

We know that the fourth order linear differential operator
$L_1^s \oplus (Z_2 \cdot N_1)$ is a right-most factor of $L_{11}$, so we obtain
\begin{eqnarray}
L_{11} \, \,=\,\,\,\, N_7 \cdot (L_1^s \oplus (Z_2 \cdot N_1)).
\end{eqnarray}
The ODE formula for the seventh order linear  differential operator $N_7$ reads:
\begin{eqnarray}
N \, = \, \, \, 
15\, Q \,  + 7\, D \,  +89 \,\, \,=\,  \,\, \, (Q+1)(D+1) \, \, -f.
\end{eqnarray}
At $w=\, 0$, the local exponents (for $N_7$) are
\begin{eqnarray}
\rho \, =\,\, \, \,\, \,2^2,\,\, \, 3,\,\, \, 4^2,\,\, \, 5,\,\,\, 12.
\end{eqnarray}
Plugging the series
 $w^5 \, + \sum_{k\geq 6} a_k \cdot w^k$ into the linear ODE for $N_7$
fixes all the coefficients except $a_{12}$  corresponding to a solution with 
the highest local exponent. Letting the combination coefficient $a_{12}$ vary 
in the finite range $[1, p_r]$, we found for the value $a_{12}=\, 22292$ a linear
 ODE of order less than seven, with ODE formula
\begin{eqnarray}
N \, = \, \, \, 
13\, Q \, \,+ 5\, D\, \,+79 
\,\, \,=\,\,\, \,\,(Q+1)(D+1)\,\,-f,
\end{eqnarray}
and with exponents at the origin
\begin{eqnarray}
\rho \, =\, \,\, \, 2^2, \,  \,\,3, \,\, \, 4, \,\, \, 5.
\end{eqnarray}
\noindent
For this linear ODE we consider  the one-parameter series that starts with $w^4$
and which contains  the coefficient $a_5$  as a ``free'' parameter.
By letting the coefficient $a_5$ vary in the finite
range $[1, p_r]$, we found that for $a_5 =\, 29103$, the linear ODE
of order five breaks into two linear 
differential operators of order three
and two, $O_3 \cdot O_2$, respectively.

In conclusion we have decomposed the 
 differential operator $L_{11}$ of order 11
into six irreducible factors
\begin{eqnarray}
\label{factoL11}
L_{11}  \,=\, \, \,\, \, 
\tilde{O}_2 \cdot O_3 \cdot O_2 \cdot (L_1^s \oplus (Z_2 \cdot N_1)),
\end{eqnarray}
where the indices denote the order of the 
corresponding linear differential operators.

\subsection{The  linear differential operator $L_{11}$ in exact arithmetic}
\label{exactarith}
B. Nickel has obtained\footnote[8]{Private communication.}
 some linear differential operators that
right-divide $L_{30}$ and, especially, the
 linear differential operators
(that we call) $U_2 \cdot N_1$ and $F_3 \cdot F_2 \cdot L_1^s$.
Checking these operators against the factorization (\ref{factoL11}),
we found that $L_{11}$ has the direct-sum decomposition
\begin{eqnarray}
L_{11}  \,=\,\,\, \,
(Z_2 \cdot N_1) \,  \oplus \, 
 V_2 \, \oplus \,  (F_3 \cdot F_2 \cdot L_1^s), 
\end{eqnarray}
where $V_2$ is equivalent to $U_2$ (or to $\tilde{O}_2$)
and  the product $F_3 \cdot F_2$ is equivalent
 to the product $O_3 \cdot O_2$ in (\ref{factoL11}).

Furthermore, using some tricks in the modular calculations
supplemented with constraints on the apparent polynomials
(\cite{ince-56}, Appendix A in~\cite{bo-gu-ha-je-ma-ni-ze-08}),
Nickel succeeded in finding the considered linear differential operators
exactly. The linear differential operators $V_2$, $F_2$ and $F_3$ are
given\footnote[1]{We are grateful to B. Nickel for these results.}
in \ref{Sec:ExactOp}. 

The procedure for obtaining a 
rational number from its image modulo a prime
is known as ``{\em rational reconstruction}"
 and has many applications (for details
see e.g.~\cite{wang-ratrec,monagan-ratrec,col-enc-ratrecon}).
Consider a rational number $n/d$ 
which has the residue $u$ modulo the prime $m$.
Given $u$ and $m$, a rational reconstruction algorithm tries to recover 
the rational $n/d$ under some conditions 
on the magnitude of the unknowns
$n$ and $d$. The simple version of this condition is
\begin{eqnarray}
\label{cond-ratrecon}
2 \, N^2\, <  \,\, m,   \qquad \qquad 
N\,  = \,\,  \max\left( \vert n \vert, d\right).
\end{eqnarray}
The algorithm will then output the rational $n/d$ satisfying the
above condition, but this rational number may not be the actual one
 for the problem.
If the residue is known for several primes 
$m_i$, it is the {\em Chinese remainder}
$u$ which is considered and $m$ is the product of the primes $m_i$.

In any case, the knowledge of the order of magnitude of $n$
and $d$ is important.
For instance, the exact linear
 differential operator $F_2$ can be recovered
using the results {\em for three primes}\footnote[2]{
We have actually obtained the linear differential operator 
$L_{29}\,=\,\, L_{18} \cdot L_{11}$ for
four primes $2^{15}-19$, $\,2^{15}-49$, $\,2^{15}-51$ and $\,2^{15}-55$.}.
In our modular calculations the residues are coefficients of polynomials
occurring in Fuchsian linear ODEs.
Besides the order of magnitude, which can be guessed, the
linear ODE, once reconstructed, should satisfy certain properties.
The indicial equation should give the correct local exponents, which
have been obtained either from a linear ODE modulo a prime or from a
 diff-Pad\'e analysis~\cite{bo-gu-ha-je-ma-ni-ze-08}. For 
the apparent singularities the
conditions on the apparent polynomial should be verified.

Consider the linear differential operator $F_2$ (see \ref{Sec:ExactOp})
\begin{eqnarray}
\label{49}
F_2 \,=\,\,\, {\cal P}_2(w) \,P_{app}(w) \cdot {\rm D}_w^2\, 
+{\cal P}_1(w) \,P_{1}(w)\cdot {\rm D}_w \, + P_0(w).  
\end{eqnarray}
The singularities are known and are roots 
of the polynomial ${\cal P}_2(w)$
(the polynomial ${\cal P}_1(w)$ contains a subset of the singularities). 
We note that $P_{app}(w)$ (the apparent polynomial) and $P_1(w)$ can be
reconstructed  with two primes while $P_0(w)$ demands three
primes\footnote[3]{We have in each case one or more additional
results modulo a prime for checking our calculations.}.

However, when $P_{app}(w)$ and $P_1(w)$ 
have been found, the matching of the known local
exponents will fix some coefficients in $P_0(w)$, but, more importantly, we
get an idea about the order of magnitude of the common denominator
in the various rational coefficients in $P_0(w)$. This magnitude was found 
to be $2^{16}$ and, with this scaling, the still unreconstructed coefficients in 
$P_0(w)$ can be reconstructed using two primes  
and then checked against the conditions for the apparent singularities. 
Our iterative procedure for reconstructing the exact polynomials thus amounts 
to first reconstructing the polynomials $P_{app}(w)$ and $P_1(w)$ (with
two primes) in order to obtain $P_0(w)$ with two primes instead of three.
More details can be found in \ref{Sec:Experiment}, which  
deals with a rational reconstruction experiment on the
apparent polynomial of $F_3$. One should note 
that, to rationally reconstruct
a linear ODE, it must be obtained with as many primes as necessary.
When the results are time consuming or hard to obtain, the knowledge of
the underlying problem may help one to guess 
the scaling factor which forces
the condition (\ref{cond-ratrecon}).

Note that when the number of primes  is not sufficient some reconstructed
coefficients will, obviously, be in error. A strong check can finally be done 
on these linear differential operators that should convince one of their correctness.
The linear differential operators $F_2$ and $F_3$, that we are looking for in exact
arithmetic, are factors of the linear differential operator $L_{33}$. They are 
{\em necessarily globally nilpotent}~\cite{bo-bo-ha-ma-we-ze-09} since $L_{33}$ is.
The linear differential operator $L_{33}$ is globally nilpotent since it corresponds
to a linear ODE that annihilates an integral of an algebraic integrand (\ref{chi3tild}) 
(it is {\em ``derived from geometry''},  DFG, see~\cite{bo-bo-ha-ma-we-ze-09}
and references therein). We have calculated the $p$-curvatures of these reconstructed 
linear differential operators $F_2$ and $F_3$, and found that they are, indeed,
globally nilpotent.

The global nilpotence of a linear differential operator is a strong and
{\em very special} property that is rigid enough to make us {\em totally confident}
that the polynomials occurring in the linear differential operators $F_2$ and $F_3$
have been reconstructed correctly. 

As for the solutions of the factors occurring in $L_{11}$, 
the simple $V_2$ is equivalent 
to the linear differential operator $L_E$ and its linear ODE annihilates
\begin{eqnarray}
\label{V2sol}
{\frac{w^2}{1-4w}} \cdot  \Bigl( K\,  - {\frac{2\,E}{1-16w^2}} \Bigr).
\end{eqnarray}
{}From (\ref{V2sol}) it is straightforward to see that   
$V_2$ is actually equivalent to the second order operator 
for $\, \tilde{\chi}^{(2)}$ (denoted $\, N_0$ 
 in~\cite{ze-bo-ha-ma-05b}).
Note that $V_2$, equivalent to $N_0$, does 
{\em not} mean that $\, \tilde{\chi}^{(2)}$ itself
is a solution of $L_{29}$, but 
rather some linear combination of  $\, \tilde{\chi}^{(2)}$  
and its first derivative is. Indeed  (\ref{V2sol}) can be expressed as:
\begin{eqnarray}
\frac{(1+4w)(1+8w^2)}{2w} \cdot 
\frac{\rmd \, \tilde{\chi}^{(2)} }{\rmd w}  \, \, \, \, \, 
-4 \cdot (1+4w)\cdot   \tilde{\chi}^{(2)}.
\nonumber 
\end{eqnarray}
The remarkably simple result (\ref{V2sol}) begs the question about the very 
nature of $\, V_2$.  Is the occurrence of $\, \tilde{\chi}^{(2)}$ (and its  first derivative) 
in $\, \tilde{\chi}^{(5)}$ a mere coincidence or does it suggest a more general structure.
Does an operator equivalent to $\, \tilde{\chi}^{(4)}$ appear in $\, \tilde{\chi}^{(7)}$
(or $\, \tilde{\chi}^{(3)}$ in $\, \tilde{\chi}^{(6)}$)? If we do not have this very strong 
result, is it nevertheless the case, that some of the individual factors occurring in say
the factorizations of $L_7$ appears in $\, \tilde{\chi}^{(6)}$? Expressed more
generally is it the case that operators
 equivalent to factors from  $\, \tilde{\chi}^{(m)}$
appear in the factorization of $\, \tilde{\chi}^{(n)}$ (with $m\leq n$ and $n$ and $m$
of different parity)? 
Similar questions can be asked with regard to the occurrence of the rational solution 
of $L_1^s$ that we have written as 
$\, \left(\tilde{\chi}^{(1)}\right)^2$ in (\ref{chi1square}).
We have already conjectured in (\ref{conject1}) that for the $\, \tilde{\chi}^{(n)}$'s
we have direct-sum structures corresponding to linear ODE's of smaller order 
for selected linear combinations involving $\, \tilde{\chi}^{(n-2k)}$. Could it be 
that polynomial (i.e. non-linear) combinations
 of $\, \tilde{\chi}^{(m)}$'s  ($m<n$) can be
used to further deplete $\, \tilde{\chi}^{(n)}$ or at least appear as factors?
   
As emphasised in~\cite{bo-bo-ha-ma-we-ze-09}, we have a strong belief
(but no proof) that all the linear differential operators occurring as factors 
of the  linear differential operators for the $\, \tilde{\chi}^{(n)}$s,
have to {\em be related to the theory of elliptic curves} 
(complete elliptic integrals
$\, E$ and $\, K$, algebraic modular functions expressed as algebraic 
hypergeometric functions, {\em modular forms of weight-1}, etc.).
The calculation of the $\, p$-curvature of  $F_2$ and $F_3$ excludes 
linear differential operators associated with algebraic functions.
The simple occurrence in (\ref{V2sol})
 of $\, E$ and $\, K$   is in contrast 
to the linear differential operators $F_2$ and $F_3$,
which do not seem to be equivalent to a symmetric power of $L_E$. 
We must explore whether (similarly to what 
we found~\cite{bo-bo-ha-ma-we-ze-09}  for  $\, Z_2$ 
and the linear differential operator occurring
in the analysis\footnote[9]{The functions $\Phi_H^{(n)}$
are simplified Ising type integrals~\cite{bo-ha-ma-ze-07b} obtained 
from the $\tilde{\chi}^{(n)}$ integral 
representation (\ref{chi3tild}) 
by setting the Fermionic factor $G^{(n)}=\, 1$.} of 
$\, \Phi_H^{(3)}$)
the linear differential operators  $F_2$  (resp. $F_3$)
correspond to modular forms of weight-1 (or higher weights), or $\, _2F_1$
(resp. $\, _3F_2$) hypergeometric functions with a Hauptmodul
pull-back (up to multiplication by some
 $\,n$-th root of a rational function).

To see if a solution of the second order operator $\, F_2$ is a  $\, _2F_1$ 
hypergeometric function with a Hauptmodul pull-back (up to multiplication by
the $\,n$-th root of some rational function) would require one to
find not only  the Hauptmodul pull-back, but also a change of variable (covering) 
mapping the large set of singularities in $\, F_2$ onto three singularities
($0$, $\, 1$, $\, \infty$), and find, besides, the rational function
occurring in the multiplicative factor in front of the hypergeometric function 
$\, _2F_1$ (which looks like the Hauptmodul~\cite{bo-bo-ha-ma-we-ze-09}). 

The occurrence of an involved apparent polynomial is a quite severe
obstruction for performing these educated guesses.
It is always possible to get rid of the apparent polynomial
of a linear differential operator
by introducing  a higher order, but still Fuchsian, linear
differential operator with no apparent singularities (see \ref{Sec:ExactOp}).
This however is not helpful. What we really need is to find an equivalent
linear differential operator with a smaller 
apparent polynomial or, hopefully, 
 no apparent polynomial.

This is how we achieved~\cite{bo-bo-ha-ma-we-ze-09} such 
a calculation for the linear differential operator~$Z_2$.
We were able to find a second order operator (occurring as a factor
 in  $\, \Phi_H^{(3)}$), which is simpler than $\, Z_2$
because its apparent polynomial is just $\, 1\, -2w$
\begin{eqnarray}
\label{equiv}
Z_2 \cdot M_1 \, \,  = \, \, \,\,   \tilde{M}_1 \cdot \tilde{Z}_2,
\end{eqnarray}
where $\, M_1$ and $\, \tilde{M}_1$ are  two first order
 linear differential intertwiners. 
Up to the change of variable (covering) 
\begin{eqnarray}
\label{covering}
x \, = \,  \, \,{{ 72\,  w} \over {(1\, -w) \, (1\,-4\,w)}},
\end{eqnarray}
wrapping the seven singularities of $Z_2$ onto the three singularities
of the hypergeometric function, we were able to find a 
{\em modular form of weight-1} solution of the 
equivalent second order operator 
$\,\tilde{Z}_2$.  This was a consequence of 
its very simple apparent polynomial.
At the present moment, we have not been able to replace 
$\, F_2$ by an equivalent second order operator with 
a simpler apparent polynomial. 
The situation is even more involved for $\, F_3$ (see \ref{Sec:ExactOp}).
The modular form interpretation of $\, F_2$ and $\, F_3$ remains 
to be done and is clearly a worthy challenge.

\subsection{The linear differential operator $L_{13}$}
\label{fourfactor}
We continue our procedure of factorization for the 13th order operator
$L_{13}$ occurring as a factor in $L_{29}$.
Recalling the local exponents at $w=0$
\begin{eqnarray}
\rho \, =\,\,\,  \,
1^2,\,\, \, 2,\,\, \, 3,\,\, \, 4,\,\,  \,5^2,\,\, \, 6^2,\,\, \, 7^2,\,\, \, 8,\,\, \, 9
\end{eqnarray}
we know that the series corresponding to the highest exponent enables one to
reconstruct the full linear ODE. The next series to consider is thus the
one-parameter series $w^8 +a_9 \, w^9 +\cdots$.
For every value of the linear combination coefficient $a_9$ in the
interval $[1, p_r]$, we found that the resulting linear ODE is of
order thirteen.

Both series ($w^9\,  + \cdots $ and $w^8 \, + a_9\,w^9 \, +\cdots$)
 are annihilated 
by a left-most factor in $L_{13}$. To  proceed with the factorization, we would 
have to consider ``deeper" combinations of series solutions
(see Section~\ref{Sec:FacExp}).
For instance at $w=\, 0$, we could use the two-parameter ($a_8,\, a_9$) solution
$w^7 \, + a_8\, w^8 \, + a_9\, w^9\,  +\cdots$,
then the three-parameter ($a_7,\, a_8,\, a_9$) solution
$w^6\, +a_7\,w^7\,  + a_8\, w^8\,  + a_9\, w^9\,  +\cdots$, and
finally the four-parameter solution
$w^5\, +a_6\,w^6\, +a_7\,w^7\,  + a_8\, w^8\,  + a_9\, w^9 +\cdots$. However,
if $t_0$ is the computational time for a  single ODE search,
then to check the factorization using a solution with $k$ free parameters
requires a computational time of $t_0\, p_r^k$. This requires
 a very long time for the  prime $p_r=\, 32749$ taking into account
 the sizes of the linear ODEs we are dealing with here, and 
hence we have not pursued this approach beyond the one-parameter case.
We could also use, from the outset, the most general five-parameter solution
$a_5\, w^5 \, +a_6\,w^6 \, +a_7\,w^7 \, + a_8\, w^8\,  + a_9\, w^9\,  +\cdots$,
(only $a_5=\, 0$ and $a_5=\, 1$ need to be considered) that should give all 
the factorizations (if any) of $L_{13}$.
However, a check of the factorization using this  five-parameter series
solution clearly suffers from the prohibitive time requirements mentioned 
above and it is beyond our current computational resources.

In Section~\ref{Sec:FacExp} we described our method of factorization modulo a
prime by focusing on the singularity at the origin.
This singular point has no special properties which makes it better suited than
other singular points for our factorization scheme.
However, to have a clear working scheme, the singular point one chooses to
focus on must be sufficiently singular (by which we mean, in this case, have
many confluent logarithms) to allow one to extend from
the {\it local scheme} to the {\it global scheme}.
So, we looked at what happens if we use expansions about other singular points.

Considering the ODE corresponding to $L_{13}$ 
translated to\footnote[5]{The local exponents at $w=\, \infty$ are
$-40, -39^2, -38, -36^2, -35, -34^2, -33, -32, -31, -30$.} $w=\infty$
one can follow the series of the highest exponent which is $-30$.
This series also demands the full ODE. The one-parameter series corresponding
to the second highest exponent
 $x^{-31}\,  + a_{-30}\, x^{-30}\, +\cdots$, (with $x=\, 1/w$)
also gives rise to the full ODE (i.e. the order remains 13)
for all values of $a_{-30} \in [1, p_r]$.
Similar calculations were performed for the linear
 ODE translated to\footnote[8]{The local exponents at $w=1/4$ are
$-9, -8, -7, -6, -5, -4, -3, 0, 1, 2, 3, 4, 5$.} $w=1/4$.
Neither the series $x^5+\cdots$, nor the one-parameter series
 $x^4\, +a_5\, x^5\,  +\cdots$,
(with $x=\, w-1/4$) gives rise to a linear ODE of order less than thirteen
for any value of $a_{5} \in [1, p_r]$. 
The series solutions in front of the higher logarithmic
solutions around some other
singular points give rise to the full linear ODE
of order thirteen.

Instead of considering the series solution 
$a_5\, w^5\, +a_6\,w^6\, +a_7\,w^7\,  + a_8\, w^8\,  + a_9\, w^9\,  +\cdots$,
with its prohibitive computational time requirements, 
we decided to try another procedure that 
may give us an idea about the number and order of factors occurring
in $L_{13}$ (if reducible).

We start by examining how the various formal solutions of $L_{13}$ appear.
Consider (near $w=0$) a general solution with log's such as
\begin{eqnarray}
S(w) \, =\,\, \,  S_n(w) \,\log(w)^n \, 
+ S_{n-1}(w) \, \log(w)^{n-1}\,  + \cdots\,  + S_0(w), 
\end{eqnarray}
where the exponent $n$ of the log is generically taken to the maximum allowed
value of 12, i.e. the order of $L_{13}$ minus one, and where the $S_n(w)$ are
power series expansions $a_0^{(n)}\,  +a_1^{(n)} \,w\,  +\cdots$.
Plugging the solution $S(w)$ into the linear 
ODE corresponding to $L_{13}$ and solving $L_{13} \left( S(w) \right)=0$
term by term, we found that the highest allowed exponent is $n=\, 3$.

We therefore fix the solution $S(w)$ as
\begin{eqnarray}
&& \fl  \qquad \quad
S(w) \, =\, \, \, S_3(w) \,\log(w)^3 + S_{2}(w) \, \log(w)^{2}\, 
\ +\,  S_{1}(w) \, \log(w)\,+ S_{0}(w), 
\end{eqnarray}
and act on it by the linear ODE corresponding to $L_{13}$ and solve term 
by term (we have to solve only to $w^9$ since this is the highest local 
exponent for $L_{13}$ around $w= \, 0$).
The coefficients (up to $w^9$) in the $S_k(w)$ must be fixed. Among
the 40 coefficients 13 will remain as free parameters 
(equal to the order of the linear ODE).
Attached to any of these free coefficients is an independent solution of
the linear ODE.
 
To clarify the scheme of these solutions, from which we shall infer
the number of factors, we solve $L_{13} \left( S(w) \right)=0$.
This leads to the equation
\begin{eqnarray}
\sum_{k \ge 0} \, \Bigl( \sum_{n=0}^3 \, C_k^{(n)}\, \log(w)^n \Bigr)  w^k  \,= \,0,
\end{eqnarray}
which we solve for each $k$ and  $n$ by using the following recipe:
The coefficient $C_k^{(n)}$ of the term $w^k \log(w)^n$ will in general be
a linear combination of coefficients $a_j^{(m)}$  from $S_m(w)$ with
$j\leq k$ and $n \leq m$.  
When the coefficient $C_k^{(n)}$ contains  coefficients  $a_j^{(m)}$ 
with  $m=n$ only, we solve for the coefficient $a_j^{(n)}$ of highest index $j$.
When the coefficient $C_k^{(n)}$ contains coefficients $a_j^{(m)}$  with  $m\leq n$,
we solve for the coefficient $a_j^{(m)}$ of lowest index $m$ and highest index $j$.
This is because we know, from all our computations on Ising type ODEs,
that when a solution such as
$S(w)\,\log(w)^{(n)} \, +\cdots$ occurs,  
$S(w)\,\log(w)^{(n-1)} \, +\cdots$, (with the
same $S(w)$) is also a solution.

We introduce the notation $[w^p]$ to mean that the series
 begins as $w^p \,(const. + \cdots)$.
The results of the computation
are the following. Four solutions can be written as
\begin{eqnarray}
\label{series7}
&& [w^7] \, \log(w)^3\, + [w^5] \, \log(w)^2 \,+
[w] \, \log(w) \, + [w],  \,  \nonumber \\
&& [w^7] \, \log(w)^2\, + [w^5] \, \log(w)\, + [w],   \nonumber \\
&& [w^7] \, \log(w) \, + [w], 
\qquad \quad \hbox{and} \quad  \qquad  [w^7]. 
\end{eqnarray}
Four other solutions can be written as
\begin{eqnarray}
\label{series6}
&& [w^6] \, \log(w)^3\, +[w^5] \, \log(w)^2 \,+
[w] \, \log(w) \,+ [w],  \nonumber \\
&& [w^6] \, \log(w)^2 \,+ [w^5] \, \log(w) \, + [w] ,
 \nonumber \\
&& [w^6] \, \log(w) \, + [w], 
\qquad   \quad  
\hbox{and} \qquad  [w^6].  
\end{eqnarray}
There are also two sets of two solutions each that read
\begin{eqnarray}
\label{series9}
 [w^9] \, \log(w) \, + [w], \qquad\quad 
\hbox{and} \qquad \quad   [w^9], 
\end{eqnarray}
and
\begin{eqnarray}
\label{series8}
 [w^8] \, \log(w) \, + [w], 
\qquad \quad  \hbox{and}\quad  \qquad  [w^8].  
\end{eqnarray}
Finally there is a non-logarithmic solution starting as $w^5+\cdots$.

In view of this scheme, one may conclude, in analogy with all the Ising
calculations we have performed and where hypergeometric functions occur,
that the factors occurring
in $L_{13}$ are of order $4,  \, 4,  \, 2,  \, 2$ and $1$.
At the point $w= \, \infty$, one obtains the same structure of solutions
leading to the same scheme, that is factors of order 
$4, \,  4, \,  2, \,  2$ and $1$. However,
at the singular point $w=1/4$ the structure changes slightly.
The solutions, grouped as done above for the point $w= \, 0$, lead to a 
scheme of six factors with orders $4, \,  2, \,  2, \,  2, \,  2$ and $1$.

To reconcile the three\footnote[1]{There are not enough logarithmic
solutions at the other singular points.}  
schemes (around $w=0$, $w=1/4$ and $w=\infty$),
the linear differential operator $L_{13}$ may
 have either four factors of orders
$4, 4, 4$ and $1$ or five factors of orders $4, 4, 2, 2$ and $1$.

Around the three singular points, the schemes allow for an order one
differential operator whose corresponding series  starts as $w^5 +\cdots$.
It happens that this first order differential operator (call it $\tilde{L}_1$) 
occurs as a right-most factor  of $L_{13} = \, L_{12} \cdot \tilde{L}_1$.

We have found that the solution of the linear ODE
 corresponding to $\tilde{L}_1$
is a simple polynomial of degree 34,
which reads (modulo the prime $p_r$)\footnote[2]{With the solution $P(w)$,
we have the coefficients for the deepest combination series
$w^5\, +a_6\,w^6\,+a_7\,w^7 \,+ a_8\, w^8\, + a_9\, w^9\, +\cdots$.}
\begin{eqnarray}
&& P(w) \, = \, \, {w}^{5}+30849\,{w}^{6}+4080\,{w}^{7} +11244\,{w}^{8}
+26721\,{w}^{9} \nonumber \\
&&\quad +29301\,{w}^{10} +23070\,{w}^{11}
+30185\,{w}^{12}+26217\,{w}^{13}+10853\,{w}^{14} \nonumber \\
&&\quad+25659\,{w}^{15}  +4536\,{w}^{16}
+31400\,{w}^{17}+22061\,{w}^{18}
+31481\,{w}^{19}\nonumber \\
&&\quad +3767\,{w}^{20} +6508\,{w}^{21}+10160\,{w}^{22}+
31426\,{w}^{23}+29441\,{w}^{24} \nonumber \\
&&\quad +17755\,{w}^{25} +6024\,{w}^{26}+31840\,
{w}^{27}+10393\,{w}^{28}+20669\,{w}^{29} \nonumber \\
&& \quad +4477\,{w}^{30} +29192\,{w}^{31}+20075\,{w}^{32}
+2957\,{w}^{33}+2003\,{w}^{34}.
\end{eqnarray}

Although we have obtained such polynomials for the {\it four primes}, previously
mentioned, our attempted rational
 reconstruction~\cite{wang-ratrec,monagan-ratrec,col-enc-ratrecon}
of the exact $P(w)$ was not successful.
There is no further information to guide
our quest for the exact $P(w)$.
There are only two indicial exponents ($5$ and $-34$) corresponding to
the two points $w=0$ and $w=\, \infty$, respectively.
The linear differential operator $\, \tilde{L}_1$
is a first order linear differential operator of the form:
\begin{eqnarray}
\label{orderone}
\tilde{L}_1 \, = \, \, \, \,
 {{\rmd} \over {\rmd w}} \, \,\,  \, + \,  {{\rmd \ln(P(w))} \over {\rmd w}}.
\end{eqnarray}
It is thus automatically globally nilpotent.
Global nilpotence is a very severe constraint
 to fulfill for higher order linear differential operators,
but for first order operators like (\ref{orderone}) it provides no additional
 constraints on $\, P(w)$.

\section{Comments and speculations}
\label{specul}
In view of the previous results, we give some concluding remarks.
The linear differential operator $L_{29}$, corresponding to 
$ \, \tilde{\chi}^{(5)}\, -\tilde{\chi}^{(3)}/2 \,
 + \tilde{\chi}^{(1)}/120$, 
 can be written as 
  \begin{eqnarray}
\label{63}
L_{29} \,=\,\, \,\, L_5 \cdot L_{13} \cdot L_{11},
\end{eqnarray}
with
\begin{eqnarray}
&&L_{11} \,=\,\,
 \left( Z_2 \cdot N_1 \right) \oplus V_2 \oplus 
\left( F_3 \cdot F_2 \cdot L_1^s \right),  \\
&&L_{13} \, = \,\, L_{12} \cdot \tilde{L}_1.
\end{eqnarray}
The linear differential operator $L_5$ is equivalent
 to the symmetric fourth power of $L_E$.
This linear  differential operator is the factor of maximum order, assuming
that the factorization scheme of $L_{12}$ is correct.
The scheme of factorization (\ref{63}) then 
generalizes what we have obtained for
$\tilde{\chi}^{(3)}$ and $\tilde{\chi}^{(4)}$.

Our conjecture on the structure of the $\tilde{\chi}^{(n)}$, namely,
(\ref{conject1}) and (\ref{conject2})
would give, for the six-particle contribution $\tilde{\chi}^{(6)}$, the
following scheme
\begin{eqnarray}
\tilde{\chi}^{(6)}
\, \,=\, \, \,\, \alpha_{4} \cdot  \tilde{\chi}^{(4)}\,  \,
+\alpha_{2} \cdot \tilde{\chi}^{(2)}\,  \, + \Phi^{(6)},
\end{eqnarray}
with $\Phi^{(6)}$ a solution of a linear ODE of order $q$ whose
corresponding linear differential operator factorizes as 
$L_{q}  \,=\,\,  L_{6} \cdot L_{q-6}$,
and where the left-factor  $ L_{6}$ is
equivalent to the symmetric fifth power of $L_E$.

As far as the singularities are concerned, the 11th order
linear differential operator $L_{11}$ has only the
 singularities of the linear ODE
corresponding to $L_7$ (the operator for $\tilde{\chi}^{(3)}$) and
the ``unknown''\footnote[3]{Unknown with respect to the
 $\Phi_H^{(5)}$ integrals~\cite{bo-ha-ma-ze-07b} and to our Landau 
singularity analysis~\cite{bo-gu-ha-je-ma-ni-ze-08}.} 
 $w= \, 1/2$. This singularity $w= \, 1/2$ 
occurs in the third order linear differential operator $F_3$.
The second order differential operator $F_2$ 
is {\em responsible for the} $\rho=\, -5/4$,
$\rho= \, -7/4$ {\em singular behavior} around the
 ferromagnetic point $w=\, 1/4$
(see Table 4 in~\cite{bo-gu-ha-je-ma-ni-ze-08}).

The singularities of the linear ODE corresponding to the block $L_{12}$
are (besides $w=\, 0, \,\pm 1/4$): 
\begin{eqnarray}
\label{singchi5}
&&  (1-w) (1+2w) (1+3w+4w^2) \nonumber \\
&& (1+w)(1-3w+w^2)(1+2w-4w^2)(1 -w -3w^2 +4w^3) \nonumber \\
&& (1 +8w +20w^2 +15w^3 +4w^4)(1-7w+5w^2-4w^3) \nonumber \\
&& (1+4w+8w^2).
\end{eqnarray}
The singularities in the first line are also singularities of the
linear  ODE
for $\tilde{\chi}^{(3)}$. 
Note that at $w=\, -1$, $w=\,1$ and $w=\,-1/2$, for instance, 
the linear differential equation corresponding to $L_{12}$ has logarithmic
solutions. Therefore, at least one of the factors 
(if the scheme is correct) in the
block $L_{12}$ is not equivalent to a symmetric power of $L_E$.
If we consider the possibility that the  linear differential
operator of order twelve $\, L_{12}$ is irreducible, 
this would mean that we are faced with a highly restricted object,
which is globally nilpotent and not 
equivalent\footnote[9]{But might, for instance, be equivalent to 
a symmetric  power of a smaller order globally nilpotent
 operator related to modular forms.}  
to the symmetric eleventh  power of $L_E$.

Let us close with the following question arising from some of the modular
calculations and rational reconstructions presented in this paper.
Is it in general easier to generate series for many primes, use these to
reconstruct the exact series and  hence obtain the exact linear ODE, or
 is it easier to obtain the linear ODE for a
smaller number of primes and then carry out the rational reconstruction on the
coefficients of the linear ODE?
To disabuse the reader of the obvious first impression that the second method 
must be easier, we would like to point out that when we opt for a non-minimal 
order linear ODE, we gain by way of a reduction in the number of terms 
necessary to find the linear ODE, but this 
comes at a cost of an increase in the size 
of the coefficients in the linear ODE
(see the last paragraph of Appendix A in~\cite{bo-gu-ha-je-ma-ni-ze-08} for
an estimate for $\tilde{\chi}^{(3)}$).
Even if we are dealing with the minimal order linear 
ODE, the coefficients in the
right-factors have less digits than the coefficients occurring in the
left-factors. For instance, considering the first factorization in
(\ref{factoL7}), the maximum number of digits is 6
 in $Z_2$, it increases to
27 for $Y_3$ and to 33 for $M_1$.

\section{Conclusion}
\label{conclusion}

Using the Fuchsian linear ODE of  $\, \tilde{\chi}^{(5)}$ (obtained modulo a 
single prime $p_r$), we have been able to go quite a long way towards 
understanding the factorization of its corresponding (minimal order) linear
differential operator $L_{33}$.
In particular we have found several quite remarkable results.

The direct-sum structure  of $L_{33}$  generalizes what we have found for 
the linear differential operators of $\, \tilde{\chi}^{(3)}$ and $\, \tilde{\chi}^{(4)}$. 
In the linear differential operators of  $\, \tilde{\chi}^{(5)}$ we found not only the 
occurrence of a term proportional to $\, \tilde{\chi}^{(3)}$, but  also the occurrence 
of a term proportional to $\, \tilde{\chi}^{(1)}$. We conjecture that this structure 
occurs for all  $\, \tilde{\chi}^{(n)}$, i.e. we 
expect to see terms in $\, \tilde{\chi}^{(n)}$ 
proportional to $\, \tilde{\chi}^{(n-2k)}$.

The  linear differential operator $L_{29}$ annihiltating the  ``depleted"  linear 
combination (\ref{combinat})  for 
$\, \tilde{\chi}^{(5)}$ follows the same structure 
seen for $\, \tilde{\chi}^{(3)}$ and $\, \tilde{\chi}^{(4)}$.
The left-most factor of $L_{29}$ is equivalent to the symmetric fourth
power of the second  order operator corresponding to
complete elliptic integrals of the first (or second) kind.

Some right-factors of $L_{29}$ are 
given in exact arithmetic. In particular one notes the occurrence
of a very simple second order operator $\, V_2$
and of the remarkable factor $\, Z_2$ that occurred 
in $\,\tilde{\chi}^{(3)}$. 
Using the series
of $\tilde{\chi}^{(5)}$ obtained for $p_r$ 
 as well as three additional primes, we have
obtained the linear ODE modulo each prime and checked that the mentioned
right factors are indeed exact.
We have used the results for the four primes to see whether
a rational reconstruction of right factors is feasible.

Two of the right-factors, $F_2$ and $F_3$ (of 
order two  and three, respectively), are 
highly restricted {\em globally nilpotent}
linear differential  operators, but, unfortunately,
we have not been able to find exact solutions as we did for the linear 
differential operator $Z_2$
occurring in the factorization of the linear 
differential operator for $\, \tilde{\chi}^{(3)}$.
Providing a better understanding of these operators, say
in terms of {\em modular forms}, is clearly a natural
(but actually quite difficult) challenge.

The incomplete part of our analysis is concerned with the 13th order
linear differential operator $L_{13}$.  For this operator we did find
a right-factor of order one which, {\em quite
 remarkably, has a polynomial solution}. The factorization 
of the remaining 12th order linear differential $L_{12}$
is beyond our current computional ressources. By producing 
all the twelve formal solutions of $L_{12}$, a factorization scheme appears
where the differential operator $L_{12}$ (if reducible) could factorize
into three fourth order operators, with a possible scenario 
that one of the fourth order operators could factor into two second
order operators. Clearly some work remains to be done to better 
understand  $\, L_{12}$, and hopefully find the exact fourth order
 operators in its factorization. We thus 
hope to gain a better understanding of
 their very nature (are they symmetric powers of $\, L_E$ or perhaps
 linear ODEs associated to modular forms, 
namely hypergeometric functions with
 a Hauptmodul pull-back).

\ack
We would like to thank B. Nickel for many remarks and exchange of mail
dating back to our previous paper~\cite{bo-gu-ha-je-ma-ni-ze-08} 
and for remarkably obtaining the
differential operators $F_2$ and $F_3$ in exact arithmetics.
SH and IJ thank the warm hospitality of LPTMC, where part of this
work has been completed.  AB wishes to acknowledge financial
support from the joint Inria-Microsoft Research Centre. AJG and IJ gratefully 
acknowledges financial support  from the Australian Research Council. 
The calculations of the series for $\tilde{\chi}^{(5)}$  
were performed on the facilities of the Australian Partnership for  Advanced 
Computing (APAC) and the Victorian Partnership for Advanced Computing (VPAC). 
This work has been performed without any support
 of the ANR, the ERC, the MAE. 

\appendix

\section{Solution of the order four linear ODE $M_2$
occurring in $\tilde{\chi}^{(4)}$}
\label{Sec:M2Sol}
Defining $x=\, w^2$ and
\begin{eqnarray}
\fl \qquad \quad
K  \, =\, _2 F_1 \left( [1/2, 1/2], [1], 16x \right), \quad
E  \, =\, _2 F_1 \left( [1/2, -1/2], [1], 16x \right)
\end{eqnarray}
the solution (analytical at $x= \, 0$) of the linear 
ODE corresponding to $M_2$ is
\begin{eqnarray}
\fl \qquad \quad
{\frac{1}{x^4 \cdot  (1-16x)^{4} (1-4x)(7+80x)}} \cdot
\Bigl((1-16x) \cdot  P_{3,0} \cdot K^{3} \, 
 \, -3 \cdot P_{2,1} \cdot K^2\,E \nonumber \\
\fl \qquad \qquad \qquad \qquad \qquad \qquad \qquad \qquad 
 -3 \cdot  P_{1,2}\cdot  K \, E^2  \, 
-3 \cdot  P_{0,3}\cdot E^3   \Bigr)
 \nonumber
\end{eqnarray}
with:
\begin{eqnarray}
\fl \quad P_{3,0} \,  = \,  \,     819200\,{x}^{5} -1050624\,{x}^{4}
+494976\,{x}^{3}-39128\,{x}^{2}  -90\,x +63,   \nonumber \\
\fl \quad P_{2,1} \, =\,\, 
 26214400\,{x}^{6}+1458176\,{x}^{5}-4698112\,{x}^{4}
+678464\,{x}^{3} -26120\,{x}^{2}-818\,x +63,  \nonumber \\
\fl \quad P_{1,2}  \, =\, 
7208960\,{x}^{5} +1169408\,{x}^{4}
 -300288\,{x}^{3} +8728\,{x}^{2}
 + 538\,x -63,    \nonumber \\
\fl \quad P_{0,3}  \, = \, 
 363520\,{x}^{4}+53696\,{x}^{3}-1144\,{x}^{2} -86\,x +21.
  \nonumber 
\end{eqnarray}

\section{The ODE formula}
\label{Sec:ODEformula}
The linear differential equations annihilating a series $S(x)$ we are interested 
in are Fuchsian. This means that all singular points of the linear ODE, and in 
particular $x=\, 0$ and $x=\, \infty$, are regular. A form of the linear differential
operator that automatically satisfies this constraint is: 
\begin{eqnarray}
\label{LQD}
        L_{QD} \,  = \,\,  
 \sum_{i=0}^{Q} \Bigl(  \sum_{j=0}^{D}\,  a_{i j} \cdot  x^j \Bigr)
 \cdot \left( x\, {{\rmd} \over {\rmd x}} \right)^i,  \quad
 a_{Q 0}\, \ne \, 0, \,\, \, a_{Q D} \,\ne \, 0. 
\end{eqnarray}
The condition $\, a_{Q 0}\, \ne \, 0$ (resp. $\, a_{Q D}\, \ne \, 0$) is
required to make $\, x=\, 0$ (resp. $x=\infty$) a regular singular point.

Note that it is the use of the (homogeneity\footnote[5]{Also called Euler's operator.
Recall that 
$\left( x\, {d \over dx} \right)^n \cdot x^k =\, k^n\, x^k$.})  operator 
 $x\, {{\rmd} \over {\rmd x}}$ (rather than just $\rmd/\rmd x$), which leads to the 
above conditions guaranteeing the regularity of  the singular points 
$\, x=\, 0$ and $x=\infty$  and to the equality  of the degrees of the polynomials 
in front of the derivatives. For the operator $\rmd/\rmd x$, a simple rearrangement 
of terms shows that (\ref{LQD}) can be re-written as
\begin{eqnarray}
\label{LQDdx}
        L_{QD}\,  = \, \, \, \,
\sum_{i=0}^{Q}  \Bigl( \sum_{j=0}^{D} \, b_{ij} \cdot x^{j+i} \Bigr) 
\cdot \left( {{\rmd} \over {\rmd x}}\right)^i, 
\end{eqnarray}
where the coefficients $\, b_{ij}$ are linear combinations of the $ \,a_{ij}$.
This is the form of the Fuchsian linear ODE we have used in many previous papers 
(e.g.~\cite{bo-ha-ma-ze-07,bo-ha-ma-ze-07b,ze-bo-ha-ma-04,ze-bo-ha-ma-05b,Diag,bo-ha-ma-mc-or-ze-07}).
The Fuchsian character of the ODE is reflected in the decreasing degrees of the
successive polynomials in front of the derivatives.

To find the linear ODE annihilating $S(x)$, the coefficients  $a_{i j}$ in (\ref{LQD}) 
have to be determined. This can be done by demanding
$L_{QD}\left(S(x)\right)\,=\, \, 0$, resulting in a set of linear equations
for the unknown coefficients  $a_{i j}$.
In~\cite{bo-gu-ha-je-ma-ni-ze-08} this set of linear equations was put
into a  well defined order and if the corresponding  $\,N_{QD}\times N_{QD}$
determinant (with $\,\,  N_{QD}\, =\, \, (Q+1)(D+1)$) vanishes,
a non-trivial solution exists. The zero-determinant condition was checked
by creating an upper triangular matrix $U$ using standard Gaussian elimination
and a solution exists if we find $ U(N,N)=\, 0$ for some $\, N$.
The $\, N $ for which $\, U(N,N)\,=\,\, 0$ is thus the  minimum number of
coefficients needed to find the linear ODE for given  $\, Q$ and  $\, D$.
The deviation between the actual number of coefficients needed $N$, and
the generic (maximum) $\, (Q+1) (D+1)$ was called $\Delta$
in~\cite{bo-gu-ha-je-ma-ni-ze-08}.

To fully understand the deviation $\Delta=\, (Q+1) (D+1)-N$, we may
alternatively compute the nullspace of the matrix $U$.
The dimension of the nullspace, if a solution
 exists, is related to $\Delta$. In others words,
solving $L_{QD}\left(S(x)\right)\,=\, \, 0$ term by term will fix all the
coefficients but leaves $f$ coefficients unfixed among the $N_{QD}$ ones.
These are all independent ODE solutions for given $Q$ and $D$.

In~\cite{bo-gu-ha-je-ma-ni-ze-08} we reported a remarkable formula
arising from empirical observation  
\begin{eqnarray}
\label{empirical}
N \, = \, \, \,   
d \cdot Q \, \,  + q \cdot D \, \,  -C \,=  \,  \, \, (Q+1)(D+1)\, \,  -f,
\end{eqnarray}
where we have replaced the parameter $\Delta$ used in~\cite{bo-gu-ha-je-ma-ni-ze-08}
by the parameter $f$ that we now can understand {\em as the number of independent
solutions} for given $Q$ and $D$ (this understanding will be useful later).
The ODE formula (\ref{empirical}) should be understood as follows:
For a long series $S(x)$ we use three (or more) sets of $Q$ and $D$ and
solve $L_{QD}\left(S(x)\right)\,=\, \, 0$
 (by nullspace or term by term). From this
we obtain the value of the parameter $f$ (if $f>0$, otherwise we increase
$Q$ and/or $D$) for each pair $(Q,D)$.
These values ($Q$, $D$, $f$) are then used to 
determine $d$, $q$ and $C$ in (\ref{empirical}).
In all cases we have investigated, the parameter $q$ is the order of the
minimal order linear ODE that annihilates  $S(x)$.
The parameter $d$ is the number of singularities (counted with multiplicity) 
excluding any apparent singularities and the ``true'' singular point $x=0$ 
which is already taken care of  by the use of the  differential 
operator $x {\rmd \over \rmd x}$.

We revisit in Table \ref{Ta:1} some ODE formulae from
Table~1 of~\cite{bo-gu-ha-je-ma-ni-ze-08}. We give 
the value of the parameter $f$
corresponding to the same $Q$ and $D$ 
considered in~\cite{bo-gu-ha-je-ma-ni-ze-08}.
The first observation is that, generally, both
 ODE formulae (in Table \ref{Ta:1} and
 in~\cite{bo-gu-ha-je-ma-ni-ze-08}) agree.
When they do not, the difference
is in the parameter $C$. But we remark that 
$C-f$ always equals $C\, -\Delta$, which is
 easily understood from the equality
in (\ref{empirical}).
The second observation is that, for the linear 
ODE which have the constant as
solution  (i.e. $\tilde{\chi}^{(4)}$ 
and  $6\tilde{\chi}^{(4)}\, -2\tilde{\chi}^{(2)}$),
the parameter $q$ appears as the actual one.

\begin{table}[htdp]
\caption{
 ODE formula for $\tilde{\chi}^{(n)}$, $n=\,1,\,2,\, \cdots,\, 5$ and for the
 combinations $6\, \tilde{\chi}^{(n+2)}\,-n \, \tilde{\chi}^{(n)}$,
 $n=\, 1, \,2,\, 3$.
The last column gives the value of the parameter $f$
corresponding to the same $Q$ and $D$ 
considered in~\cite{bo-gu-ha-je-ma-ni-ze-08}.
 }
\label{Ta:1}
\begin{center}
\begin{tabular}{|c|c|c|c|c|}\hline
    Series &  $d \, Q + q \,D -C $ &  $Q$ &   $D$ &      $f$ \\ \hline 
\hline
    $\tilde{\chi}^{(1)}$ &$1\, Q \, + \, 1\, D \, + 1$ &   1  &  1&  1 \\
    $\tilde{\chi}^{(2)}$& $1\, Q \, + \, 2\, D \, + 1$ &  2  &  1&   1 \\
    $\tilde{\chi}^{(3)}$& $12\, Q \, + \, 7\, D \, -37$ &  11  &  17&   2 \\
    $\tilde{\chi}^{(4)}$& $7\, Q \, + \, 10\, D \, -36$ &  15  &  9&  1 \\
    $\tilde{\chi}^{(5)}$& $72\, Q \, + \, 33\, D \,-887$  & 56 & 129  &  8  \\
    $6\tilde{\chi}^{(3)}-\tilde{\chi}^{(1)}$&  $12\, Q \, + \, 6\, D \, -26$ & 10 & 17 &   2  \\
    $6\tilde{\chi}^{(4)}-2\tilde{\chi}^{(2)}$&  $6\, Q \, + \, 8\, D \, -17$ & 13 & 8 &   1  \\
    $6\tilde{\chi}^{(5)}-3\tilde{\chi}^{(3)}$& $68\, Q \, + \, 30\, D \,-732$  & 52 & 120 &  9 \\
\hline
 \end{tabular}
\end{center}
\end{table}

With the nullspace computation we now understand the constant $f$ 
($\Delta_0$ in~\cite{bo-gu-ha-je-ma-ni-ze-08}). Thus for the minimal 
order ODE, one should have $f=\, 1$ since the minimal order ODE is unique.
Setting  $Q\, =q$ and $D\, =\,\,  d+D_{app}$, where
$D_{app}$ is the degree of the polynomial whose roots are apparent
singularities, one obtains the {\em exact relation} 
\begin{eqnarray}
\label{Dapp}
D_{app}  \,=\, \, (d-1)(q-1)\, -C\,  -1
\end{eqnarray}
between the constant $C$ and the degree $D_{app}$. For  
$\tilde{\chi}^{(3)}$ one has $d=12$, $q= 7$ and
 $C=37$ giving $D_{app}=28$
which is~\cite{ze-bo-ha-ma-04,ze-bo-ha-ma-05} the degree of 
the polynomial carrying apparent singularities
in the linear ODE of $\tilde{\chi}^{(3)}$.
For $\tilde{\chi}^{(4)}$ one has $d=7$, $q=10$ and 
$C=36$ giving $D_{app}=17$,
which is~\cite{ze-bo-ha-ma-05b} the degree of the polynomial 
carrying apparent singularities
in the ODE of $\tilde{\chi}^{(4)}$.
Similarly for $\tilde{\chi}^{(5)}$, with
 $d=72$, $q=33$ and $C=887$ we
obtain the degree of the apparent polynomial $D_{app}=1384$,
which is in agreement with what appears 
in the linear ODE for $\tilde{\chi}^{(5)}$
reduced to its minimal order.

Note also that (\ref{Dapp}) is valid for linear ODEs without an apparent
polynomial ($\tilde{\chi}^{(1)}$ and $\tilde{\chi}^{(2)}$). But there are
cases where the parameter $C$ is negative
 while the linear ODE has an apparent
polynomial. This is the case we consider now.

\subsection{The ODE formula for the factors}
\label{ODEformfactors}
We first show how the apparent polynomials occur in a factorization 
 of linear differential operators such as
\begin{eqnarray}
 {\cal L} \, =\,\,\, \, L \cdot R \nonumber 
\end{eqnarray}
where the factors $L$ and $R$ are monic\footnote[8]{Normalization 
of the head polynomial of the linear differential operator.} and of 
minimal order, denoted respectively $q_L$ and $q_R$.
Denoting by $P_{app}$, the apparent polynomial 
occurring in ${\cal L}$, one knows
that this polynomial should appear as an apparent
 polynomial in the left-operator~$L$.
It may happen that the right operator $R$ also contains a polynomial $Q$
of apparent singularities and this
 polynomial {\em should not appear in} ${\cal L}$.
For this to happen, the left operator $L$ must have the roots of $Q$ as
true singularities. Furthermore, $Q$ should occur in $L$ to
the power of the order of $L$, i.e. as $Q^{q_L}$. 
The local exponents for $L$ at any root of~$Q$ are
 $-1,\,  1, \, 2,\,  \cdots,\,  q_L-1$.
If we remove the singularity $Q^{-1}$ from $L$, the new linear differential
operator $\tilde{L}$ will have $Q$ as an
{\em apparent polynomial} and will occur as
$Q^{q_L-1}$ with local exponents $0,\,  2, \, 3, \, \cdots, \, q_L$.

Consider, as an example, the series $S$ 
for $\tilde{\chi}^{(3)}$ annihilated
by a seventh order linear ODE with $L_7$ as the corresponding linear 
differential operator.
We know that this operator factorizes as (among other
factorizations (\ref{factoL7}))
\begin{eqnarray}
 L_7 \, =\, \,  L \cdot R \,=\, \,\,
 (M_1 \cdot Y_3)  \cdot (Z_2 \cdot N_1).
\end{eqnarray}
Assume that the right-operator $R$ is known. The aim is to produce the
left-operator $L$ by acting on $S$ with $R$.
The series $R(S)$ will satisfy a linear ODE corresponding to $L$.

For the right-operator $R\, = \, Z_2 \cdot N_1$,
 the left-hand side 
of the ODE formula  (\ref{empirical})  reads
\begin{eqnarray}
d_R \cdot Q\,\, + q_R \cdot D \,\, -C_R  \,\, =  \,\,\,\,\,
8 \, Q \, + 3 \,D \, -9. 
\end{eqnarray}
Putting these values into (\ref{Dapp}) we obtain $D_{app}^R=\, 4$
 which is the degree
of the apparent polynomial $Q$ occurring in $R=\,Z_2 \cdot N_1$.

The linear ODE for the left-operator $L$ produced 
from the series $R \left( S \right)$
when $R$ is taken {\it monic and of minimal order},
 has the ODE formula
\begin{eqnarray}
d_L \cdot Q\,\, + q_L \cdot D\,\,  -C_L  \,\, =   \,\,\,\,\,
15 \, Q \,+ 4 \,D\, -1. 
\end{eqnarray}
The degree of the apparent polynomial for $L=M_1 \cdot Y_3$, computed by
(\ref{Dapp}), is $D_{app}^L=\,40$, which is the degree of $P_{app}$
(the apparent 
polynomial of $\, L_7$, see above) plus three
times the degree $D_{app}^R$, and we still 
have the roots of $Q$ appearing
with multiplicity one in $d_L=\, 15$.

In computations modulo a prime, and for  high order linear ODEs, it is
obvious that it is easier to work with non-monic operators.
This results in removing the pole part of the polynomial $Q$, leaving its
apparent part in the left-operator $L$.

As an example, we will reproduce the series $R \left( S \right)$ with $R$ 
non-monic but still of minimal order. The left-hand side 
of the ODE formula  (\ref{empirical}),
corresponding to $L$, reads:
\begin{eqnarray}
d_L \cdot  Q\,\, + q_L \cdot D \,\, -C_L  \,\, =  \,\,\,
4 \, Q + 4 \,D\, +32. 
\end{eqnarray}
{}From (\ref{Dapp}) we obtain the degree of the apparent singularities in
$L$ as $40=\,28\,+3 \times 4$.
Furthermore, recalling the left-hand side of the ODE formula 
 (\ref{empirical}) (see Table \ref{Ta:1}) 
\begin{eqnarray}
d \, Q + q \,D -C \,\, =  \,\,
12 \, Q + 7 \,D -37
\end{eqnarray}
for the full $L_7$, one has $d=\, d_R\, +d_L$ and
\begin{eqnarray}
&&C \,\,  = \, \, \, \, C_L\, \,  + {\frac{q-q_R-1}{q_R-1}} \cdot  C_R \\
&& \qquad \quad   + {\frac{q_R}{q_R-1}} \cdot 
\Bigl( (q -q_R -1) \, D_{app}^R
 +dq_R -2q_R+q-d \Bigr). \nonumber 
\end{eqnarray}
\vskip .1cm 
{\bf Remark:} Even if the various parameters in the ODE formula are now
understood, we should recognize that 
we still do not know how this ODE
formula can be proved, nor where it comes from.
The various formulae dealing with the apparent polynomial degree
(in fact upper bounds, e.g.~\cite{kita-87,saeki-90,yoshida-97})
in Fuchsian linear ODEs introduce ingredients that go beyond our
experimental mathematics framework.

\section{Some linear differential operators in exact arithmetic}
\label{Sec:ExactOp}
The linear differential operators $V_2$, $F_2$ and $F_3$ occurring in the
decomposition of $L_{11}$ 
\begin{eqnarray}
L_{11} \, =\, \,\,  \,\,  
(Z_2 \cdot N_1) \,  \oplus  \, V_2 \, \oplus \, (F_3 \cdot F_2 \cdot L_1^s), 
\end{eqnarray}
read respectively
\begin{eqnarray}
\fl \qquad
V_2 \, =\, \,\,
 {\rm D}_w^{2}\, 
-{\frac { \left( 3+8\,w+16\,{w}^{2} \right) } {
 \, (1+4\,w)  \, (1- 4\,w) \, w }}\cdot  {\rm D}_w\,\,\, 
+4\,{\frac {1 +7\,w +  4\,w^2}{ (1- 4\,w) 
 \, (1+ 4\,w)^{2} \, w^{2}}}, 
\end{eqnarray}
and
\begin{eqnarray}
F_2 \, =\, \, \, \, \,
{\rm D}_w^2 \,\,\, \,  - {\frac{P_1}{P_2}}\cdot  {\rm D}_w \,\,\, \, 
 - {\frac{P_0}{P_2}}, 
\end{eqnarray}
with\footnote[1]{Note that the factors $(1+2w)$ and $(1-w)$ appear to the power one
in both $P_2$ and $P_1$. Linear differential operators can be Fuchsian without
having descending powers of the factors giving rise to the singularities.}:
\begin{eqnarray}
\fl \quad P_2 \, = \, \,  (1-4\,w) \cdot p_2,  
 \nonumber \\
\fl \quad p_2 \, = \, \,w \cdot  (1-4\,w) \, (1+4\,w) \, (1-w)
 \, (1+2\,w)\, (1+3\,w+4\,{w}^2) \nonumber \\
\fl \quad \qquad (1+w-24\,{w}^{2}-145\,{w}^{3}-192\,{w}^{4}
+96\,{w}^{5}+128\,{w}^{7}),  \nonumber \\
\fl \quad P_1 =
 (1-w)   (1- 4\,w)  (1+2\,w) 
 \Bigl( 40960\,{w}^{11}+24576\,{w}^{10}+51712\,{w}^{9}
 -66816\,{w}^{8} \nonumber \\
\fl  \qquad - 138176\,{w}^{7}-88704\,{w}^{6}
-29940\,{w}^{5}-5394\,{w}^{4} -272\,{w}^{3}
+92\,{w}^{2}+11\,w+1 \Bigr),  \nonumber \\
\fl \quad P_0 = 262144\,{w}^{13}-65536\,{w}^{12}
+335872\,{w}^{11}-934912\,{w}^{10}
-743424\,{w}^{9}  +703488\,{w}^{8} \nonumber \\
\fl  \qquad  +867776\,{w}^{7}+371848\,{w}^{6}+96744
\,{w}^{5}+14710\,{w}^{4} +2144\,{w}^{3}+398\,{w}^{2}+9\,w-11. \nonumber
\end{eqnarray}
It is possible to get rid of the apparent singularities occurring in $F_2$, 
 by multiplying  $F_2$ at the left, by a first order
 linear differential operator ${\cal L}_1$,
\begin{eqnarray}
\label{z1}
&& {\cal L}_1 \,=\ \,\,\,
 {\rm D}_w \,\, - {{1} \over {73326885520}} \cdot 
\Bigl( {{q_0  } \over {p_2}} \, \, 
 +{{256352914629} \over { 1 \, +3\,w \, +4\,w^2 }} \Bigr), \qquad \\
&& \hbox{with:}\qquad \qquad {{q_0  } \over {p_2}}
 \,\,  = \, \,\, \, {{\rmd} \over {\rmd w}} \ln(R(w)), 
\nonumber 
\end{eqnarray}
where $\, R(w)$ is a rational function
 (with integer coefficients) and
\begin{eqnarray}
\fl \quad q_0 \,= \, 244820905584+1372135276587\,w
+1384232623846\,{w}^{2}
-13621658367235\,{w}^{3} 
\nonumber \\
\fl \qquad \quad -150856196156313\,{w}^{4}
 -1054439469518747\,{w}^{5} -3472090747016314\,{w}^{6} \nonumber \\
\fl \qquad \quad     -3873078043825712\,{w}^{7}
+3114022565962720\,{w}^{8} 
   +12058813946690432\,{w}^{9} 
\nonumber \\
\fl \qquad \quad 
 +10882841933451520\,{w}^{10} -1075293814167552\,{w}^{11}
- 3544662480211968\,{w}^{12} 
\nonumber \\
\fl \qquad \quad -9348606615093248\,{w}^{13},
             \nonumber
\end{eqnarray}
thus yielding a third order
desingularized Fuchsian operator. This is the
 so-called ``desingularization'' procedure
which preserves the Fuchsian character of the linear differential
 operators. Note however that the desingularization procedure 
 {\em does not} preserve the
 remarkable property of global nilpotence of the
highly restricted second order differential operator $\, F_2$. The
 new desingularized  third order  differential 
operator is {\em no longer globally nilpotent} because 
 the first order  differential operator 
${\cal L}_1$ is {\em not globally nilpotent}. 
The breaking of global nilpotence comes from the factor
$\,  256352914629/(1 \, +3\,w \, +4\,w^2)$
in (\ref{z1}) which is not a logarithmic 
derivative of a rational function. 

Next we focus on the ``physical'' singularity
$\, w \, = \, \, 1/4$. One can change the operator 
$\, F_2$ into a slightly simpler one as follows:
\begin{eqnarray}
F_2 \, \quad \longrightarrow \quad  \quad  
\tilde{F}_2 \, = \, \, \, F_2 \cdot \Bigl( 1-4w \Bigr)^{-11/4}.
\end{eqnarray}
where the dot corresponds to a multiplication
 of  (differential) operators. 
It is important to note that the solutions of $\, \tilde{F}_2$ around 
the ``physical'' singularity
$\, w \, = \, \, 1/4$ are in fact Puiseux series
 in $\, u \, = \, \, (w-1/4)^{1/2}$.
In other words  $\, \tilde{F}_2$ rewritten in terms of the 
variable $\, u$ is {\em not} singular at $\, w \, = \, \, 1/4$.

The calculations performed on $\, Z_2$ yielded a modular form interpretation 
of  $\, Z_2$ (see~\cite{bo-bo-ha-ma-we-ze-09}). A  crucial step corresponded 
to discovering the covering 
\begin{eqnarray}
\label{cover}
&& w \quad \longrightarrow \qquad 
  t \, = \, \, \, {{ -8 \, w} \over {(1-w) \, (1-4\, w)}},
  \\
&& \hbox{or} \quad Q(t, \, w) \, = \, \, 0,
 \quad \hbox{with} \quad Q(t, \, w)
 \, =\,  \, 
4\, t \cdot w^2 \, - \, (5\,t-8) \cdot w \, +t \nonumber, 
\end{eqnarray}
which wraps the singularities of $\, Z_2$ onto the three singularities 
of $\, _2F_1$, namely $\, 0$, $\, 1$, $ \, \infty$. 
Do note that the apparent polynomial for $\, \tilde{Z}_2$ occurs as 
a vanishing condition of the discriminant in $\, w$ of the covering
polynomial $\, Q(t, \, w)$:
\begin{eqnarray}
\label{discrimi}
{\rm discrim}(Q(t, \, w), w) \,
 = \,\left( 9\,t-8 \right)  \left( t-8 \right)\, = \, \, 
 8\,{\frac { \left( 1-2\,w \right)^{2}}{ (1-w)  \, (1-4\,w) }}.
\end{eqnarray}
Trying to perform a similar calculation 
for $\, F_2$ in order to discover some  {\em modular
 form interpretation} for $\, F_2$, we observe that it is not 
straightforward to find a covering, such as (\ref{cover}),
 wrapping all the singularities of $\, F_2$ onto
 $\, 0$, $\, 1$ and $\, \infty$, and such that  the discriminant 
 in $\, w$ (like (\ref{discrimi})) of the corresponding covering
polynomial $\, Q(t, \, w)$, could correspond to
the quite involved apparent polynomial of $\, F_2$, namely 
$\, 1+w-24\,{w}^{2}-145\,{w}^{3}-192\,{w}^{4}+96\,{w}^{5}+128\,{w}^{7}$.
For these reasons we have failed in finding 
a modular form interpretation of the highly
 restricted linear differential operator $\, F_2$.
   
The third order linear differential operator $\, F_3$ reads: 
\begin{eqnarray}
F_3 \, = \, \,\, {\rm D}_w^3 \,
\,+ \, (1+2\,w)\, \, P_s^2 \, \, {\frac{ P_2}{P_3}} \cdot {\rm D}_w^2\,\,
  + 2\, P_s \,\,{\frac{ P_1}{P_3}}\cdot  {\rm D}_w\,\, 
 + \, {\frac{P_0}{P_3}},
\end{eqnarray}
where
\begin{eqnarray}
\fl \quad P_s \, = \, \, 
-w \cdot  (1-4\,w)  (1+4\,w) 
 \, (1+w-24\,{w}^{2}-145\,{w}^{3}
-192\,{w}^{4}+96\,{w}^{5}+128\,{w}^{7}), 
\nonumber 
\end{eqnarray}
\begin{eqnarray}
\fl \quad P_3 \,  =\,\,   
 \, (1-w)  \, (1-2\,w)  \, (1+3\,w+4\,{w}^{2})
 \, (1+2\,w)^{2} \cdot P_s^3 \cdot \,  p_3, \nonumber 
\end{eqnarray}
\begin{eqnarray}
\fl \quad p_3 =\, \,  5629499534213120\,{w}^{37}
+5348024557502464\,{w}^{36}\, 
-62874472922742784\,{w}^{35} \nonumber \\
\fl \quad  +339080589913096192\,{w}^{34}
+132348214635397120\,{w}^{33}+354600746294968320\,{w}^{32} \nonumber \\
\fl \quad  +1383732497338073088\,{w}^{31}-269118080922157056\,{w}^{30}
-1021414905992970240\,{w}^{29} \nonumber \\
\fl \quad  +401943021895024640\,{w}^{28}+
378516473892569088\,{w}^{27}-379126125978189824\,{w}^{26} \nonumber \\
\fl \quad  -181955521970962432\,{w}^{25}+182991453503356928\,{w}^{24}
+119809766351437824\,{w}^{23} \nonumber \\
\fl \quad  -34528714733649920\,{w}^{22}
-46719523456286720\,{w}^{21}-1865897472688128\,{w}^{20} \nonumber \\
\fl \quad  +9861412040736768\,{w}^{19}+1690374175916032\,{w}^{18}
-1285664678690816\,{w}^{17} \nonumber \\
\fl \quad  -304716171767808\,{w}^{16}+112170181177344\,{w}^{15}
+30517814178816\,{w}^{14} \nonumber \\
\fl \quad  -7815766123264\,{w}^{13}-2274047571904\,{w}^{12}
+456062896896\,{w}^{11} \nonumber \\
\fl \quad  +150282885872\,{w}^{10}-10690267808\,{w}^{9}
-6048942832\,{w}^{8}-486602112\,{w}^{7} \nonumber \\
\fl \quad  +33772908\,{w}^{6}+25075632\,{w}^{5}
+4670454\,{w}^{4}+13440\,{w}^{3}
-69066\,{w}^{2}-5169\,w -63,  \nonumber
\end{eqnarray}
\begin{eqnarray}
\fl \quad P_2 \,   =\,  \,  
2582544170319337226240\,{w}^{51}
+2029141848108050677760\,{w}^{50} \nonumber \\
\fl \quad -32885932997405703143424\,{w}^{49}
+193641813610004500971520\,{w}^{48} \nonumber \\
\fl \quad +20426022066743356162048\,{w}^{47}
+288714242895676430090240\,{w}^{46}  \nonumber \\
\fl \quad +618280187651267892346880\,{w}^{45}
-648919373873770257186816\,{w}^{44} \nonumber \\
\fl \quad -863129472633247214075904\,{w}^{43}
-1021011939308518347112448\,{w}^{42} \nonumber \\
\fl \quad  -220333306036159265112064\,{w}^{41}
+1659564100832816225320960\,{w}^{40} \nonumber \\
\fl \quad  +588473220873831600619520\,{w}^{39}
-1065067759683713707802624\,{w}^{38} \nonumber \\
\fl \quad  -9030793760523344150528\,{w}^{37}
+805481511795301371871232\,{w}^{36} \nonumber \\
\fl \quad  -122169749668787845595136\,{w}^{35}
-629129357422714417053696\,{w}^{34} \nonumber \\
\fl \quad  -87120833646056343339008\,{w}^{33}
+304015333576904250753024\,{w}^{32} \nonumber \\
\fl \quad  +143209349380404068483072\,{w}^{31}
-67135556652765458464768\,{w}^{30}  \nonumber \\
\fl \quad  -68161001548708224958464\,{w}^{29}
-1506006178531414900736\,{w}^{28}  \nonumber \\
\fl \quad  +15819782847593648750592\,{w}^{27}
+4086678104179764363264\,{w}^{26}  \nonumber \\
\fl \quad  -1909688698451711754240\,{w}^{25}
-970204468920909561856\,{w}^{24}    \nonumber \\
\fl \quad  +94919087350092267520\,{w}^{23}
+123918740818141650944\,{w}^{22}      \nonumber \\
\fl \quad  +4687965654930399232\,{w}^{21}
-10707547611722045440\,{w}^{20}        \nonumber \\
\fl \quad  -1144659629046790144\,{w}^{19}
+806082415949659264\,{w}^{18}          \nonumber \\
\fl \quad  +149774462467091328\,{w}^{17}
-48096268859594016\,{w}^{16}     \nonumber \\
\fl \quad  -16578560990131776\,{w}^{15}
+424243043096032\,{w}^{14} +905149437225280\,{w}^{13} \nonumber \\
\fl \quad  +139111711404072\,{w}^{12}
 -7972709043232\,{w}^{11}-6405062530332\,{w}^{10} \nonumber \\
\fl \quad  -1037367028148\,{w}^{9}+6971928216\,{w}^{8}
+30288912150\,{w}^{7} \nonumber \\
\fl \quad  +3873954375\,{w}^{6}-115755798\,{w}^{5}
-60227304\,{w}^{4} \nonumber \\
\fl \quad  -3099678\,{w}^{3} +211068\,{w}^{2}+21432\,w +315, 
\nonumber
\end{eqnarray}
\begin{eqnarray}
\fl \quad P_1\, = \, \,19267255250108152471879680\,{w}^{61}
+26483031235932970358407168\,{w}^{60} \nonumber \\
\fl \quad  -256308802040991428795957248\,{w}^{59}
+1579949167665869307621933056\,{w}^{58}   \nonumber \\
\fl \quad  +650374789771441405855531008\,{w}^{57}
+5216643706804247528946532352\,{w}^{56}   \nonumber \\
\fl \quad  +7917834014591751323461353472\,{w}^{55}
-3351835287019392824172871680\,{w}^{54}   \nonumber \\
\fl \quad  -11064588131657140234556014592\,{w}^{53}
-34985695129493606924629835776\,{w}^{52} \nonumber \\
\fl \quad  -27150264881506217380601135104\,{w}^{51}
+21916196537425570804428439552\,{w}^{50} \nonumber \\
\fl \quad  +42001979686686526227299172352\,{w}^{49}
+25840385187494624677295292416\,{w}^{48} \nonumber \\
\fl \quad  -5424492229252674644950908928\,{w}^{47}
-17794118224994570424773771264\,{w}^{46}  \nonumber \\
\fl \quad  +2915867386820035753799581696\,{w}^{45}
+7186426242807565546487283712\,{w}^{44}   \nonumber \\
\fl \quad  -14774359259974734620101967872\,{w}^{43}
-14907706789958430400446464000\,{w}^{42} \nonumber \\
\fl \quad  +8071574290338795697467293696\,{w}^{41}
+15504536229837797153841348608\,{w}^{40}  \nonumber \\
\fl \quad  +2626569595260883926907879424\,{w}^{39}
-6860129397955391680596148224\,{w}^{38}   \nonumber \\
\fl \quad  -4292305256193524038115524608\,{w}^{37}
+797725481517011621914869760\,{w}^{36}    \nonumber \\
\fl \quad  +1720922858808986948924866560\,{w}^{35}
+390533143866840910481326080\,{w}^{34}    \nonumber \\
\fl \quad  -290442221202989562927775744\,{w}^{33}
-165938014975046925940686848\,{w}^{32}     \nonumber \\
\fl \quad  +9585161397342427263533056\,{w}^{31}
+28502663270123757533921280\,{w}^{30}        \nonumber \\
\fl \quad  +4489386799471924718338048\,{w}^{29}
-2717486608897256297267200\,{w}^{28}         \nonumber \\
\fl \quad  -908343774367384075960320\,{w}^{27}
+162310791240979996000256\,{w}^{26}           \nonumber \\          
\fl \quad  +103239269328878845726720\,{w}^{25}
-7750369303138783333376\,{w}^{24}             \nonumber \\
\fl \quad  -11489552784013679223808\,{w}^{23}
-712249616867767788544\,{w}^{22}               \nonumber \\
\fl \quad  +991748945187237072640\,{w}^{21}
+252693598182584513344\,{w}^{20}                 \nonumber \\
\fl \quad  -21130273995450588928\,{w}^{19}
-20284808101979844832\,{w}^{18}                   \nonumber \\
\fl \quad  -3056348368556274592\,{w}^{17}
+345270164930943040\,{w}^{16}                      \nonumber \\
\fl \quad  +205893879174875432\,{w}^{15}
+28654368006663856\,{w}^{14}                        \nonumber \\
\fl \quad  -2030520435693824\,{w}^{13}-1374304588556840\,{w}^{12}
 -166988492206488\,{w}^{11}  \nonumber \\
\fl \quad  +12760292849076\,{w}^{10}+5484990319472\,{w}^{9}
+367504601004\,{w}^{8}            \nonumber \\
\fl \quad  -50197207920\,{w}^{7}-9218315844\,{w}^{6}
-277909095\,{w}^{5}                      \nonumber \\
\fl \quad  +48467763\,{w}^{4}+5648070\,{w}^{3}
+265293\,{w}^{2}+4620\,w-63,  \nonumber
\end{eqnarray}
\begin{eqnarray}
\fl \quad  P_0  = \, 
69634127209802640463075737600\,{w}^{70}
+102981137018052571618170896384\,{w}^{69} \nonumber \\
\fl \quad  -1033960403593443123509300559872\,{w}^{68}
+6716228494346939277472100777984\,{w}^{67}  \nonumber \\
\fl \quad  +830768383072984903026797969408\,{w}^{66}
+34119483032722461380174390755328\,{w}^{65}   \nonumber \\
\fl \quad  +37151403895216351475147854577664\,{w}^{64}
+7596402077224314128199487324160\,{w}^{63}   \nonumber \\
\fl \quad  -57748765852096741713914269532160\,{w}^{62}
-309493302673497714830630511968256\,{w}^{61}  \nonumber \\
\fl \quad  -232460008226528101141464649564160\,{w}^{60}
+23931702098177545910680337514496\,{w}^{59}  \nonumber \\
\fl \quad  +427559960442089709631493273288704\,{w}^{58}
+767305599958046596665201651613696\,{w}^{57}  \nonumber \\
\fl \quad  +238126263520324803598765665550336\,{w}^{56}
-489368606355635167460530948407296\,{w}^{55}  \nonumber \\
\fl \quad  -411394912009392610164715657625600\,{w}^{54}
-9190571673284172503536766025728\,{w}^{53}  \nonumber \\
\fl \quad  +4354675264071893610129679974400\,{w}^{52}
-197517596854575225398680040243200\,{w}^{51}  \nonumber \\
\fl \quad  -100108534915833684237428566523904\,{w}^{50}
+237797067305428725186438474235904\,{w}^{49} \nonumber \\
\fl \quad  +297736260249409824018326167224320\,{w}^{48}
+2827068864630764422668662865920\,{w}^{47}  \nonumber \\
\fl \quad  -212811801685085801466392370741248\,{w}^{46}
-132552668920641958581606566330368\,{w}^{45} \nonumber \\
\fl \quad  +36227143968974260317176152981504\,{w}^{44}
+80360530046171440422025236054016\,{w}^{43}  \nonumber \\
\fl \quad  +25044389743118121276003435151360\,{w}^{42}
-16775611588713607092922243612672\,{w}^{41}  \nonumber \\
\fl \quad  -14375299221388934261580907937792\,{w}^{40}
-797769185002542420001267122176\,{w}^{39}  \nonumber \\
\fl \quad  +3060062577366019941153762181120\,{w}^{38}
+1048552246961478552732246736896\,{w}^{37}  \nonumber \\
\fl \quad  -286300750377610217893186764800\,{w}^{36}
-238703363798670426041267257344\,{w}^{35}  \nonumber \\
\fl \quad  -6453603219285212538454671360\,{w}^{34}
+32067040600375745464846254080\,{w}^{33}  \nonumber \\
\fl \quad  +6256946928524452096094240768\,{w}^{32}
-3100756305550863462745636864\,{w}^{31}  \nonumber \\
\fl \quad  -1271279614733459684395712512\,{w}^{30}
+173570224057030875371798528\,{w}^{29}  \nonumber \\
\fl \quad  +187963836513544173604265984\,{w}^{28}
+20098454205158749911726080\,{w}^{27}  \nonumber \\
\fl \quad  -15220510128449109866076160\,{w}^{26}
-5530208120756891132317696\,{w}^{25}  \nonumber \\
\fl \quad  -31540092813892142535680\,{w}^{24}
+410046209080624124809344\,{w}^{23}  \nonumber \\
\fl \quad  +95695972411021353163264\,{w}^{22}
-3799963752408310388096\,{w}^{21}  \nonumber \\
\fl \quad  -6052638686215258044992\,{w}^{20}
-1044163538474290733536\,{w}^{19}  \nonumber \\
\fl \quad  +69113265719111269072\,{w}^{18}
+55225911443186243360\,{w}^{17}  \nonumber \\
\fl \quad  +6988584609018020432\,{w}^{16}
-714608406420145560\,{w}^{15}  \nonumber \\
\fl \quad  -313788688846958472\,{w}^{14}-23383932527942400\,{w}^{13} 
 \nonumber \\
\fl \quad  +4392065243452176\,{w}^{12}
+942992856333120\,{w}^{11}+18782060660376\,{w}^{10} \nonumber \\
\fl \quad  -11352161581890\,{w}^{9}
-1093090772088\,{w}^{8}+23284774974\,{w}^{7}  \nonumber \\
\fl \quad  +9267369222\,{w}^{6}+542276796\,{w}^{5}-59916\,{w}^{4} 
 \nonumber \\
\fl \quad  -3757362\,{w}^{3}-465618\,{w}^{2}-20622\,w -126. \nonumber 
\end{eqnarray}
\noindent
We note that $\, F_3$ can be simplified as follows:
\begin{eqnarray}
\label{C3}
&&F_3 \qquad \longrightarrow \qquad \, \, 
\tilde{F}_3 \, = \, \,  \, F_3 \cdot 
{{1} \over {\mu}}, \qquad \qquad \hbox{where:} \\
&&\mu \, = \, \, \,\,  w^2 \cdot (1\, -4w)^{9/2} \cdot 
(1\, +4w)^{7/2} \cdot (1-w) \cdot (1+2w) \nonumber  \\
&& \qquad \qquad \times 
(1+3\, w\, + \, 4\, w^2) \cdot App(F_2),  
\end{eqnarray}
where $\, App(F_2)$ denotes the apparent polynomial for $\, F_2$,
 namely $ \,1+w-24\,{w}^{2}-145\,{w}^{3}-192\,{w}^{4}
+96\,{w}^{5}+128\,{w}^{7} $, and where the dot in (\ref{C3})
denotes the multiplication of (differential) operators.  
This just amounts to multiplying the solutions
 of $\, F_3$ by $\, \mu$.
Remarkably  $\, \tilde{F}_3$ is {\em no longer singular}
at $\, w \, = \, 1$ nor at the two roots of the quadratic
 $\, 1+3\, w\, + \, 4\, w^2 \, = \, \, 0$. 
 We find the following exponents at the remaining singularities:
\begin{eqnarray}
&w \, = \, 0, & \quad \quad \quad \quad 0, \, \, 1, \, \, 3, \, \quad (\log^2), 
  \nonumber \\
&w \, = \, 1/4, & \quad \quad \quad  \quad
 0, \, \, 1, \, \, 3/2, \,    \nonumber \\
&w \, = \, -1/4, & \quad \quad \quad  \quad 
 0, \, \, 1, \, \, 5/2, \,    \nonumber \\
&w \, = \, \infty, & \quad\quad \quad   \quad -18, \, \, -18, \, \, -16, 
 \, \quad(\log^2),   \nonumber \\
&w \, = \, 1/2, & \quad \quad \quad   \quad
 0, \, \, 1, \, \, 1/2, \,    \nonumber \\
&w \, = \, -1/2, & \quad \quad \quad  \quad
  0, \, \, 1, \, \, 1/2, \,    \nonumber \\
&App(F_2) \, = \, 0, & \quad \quad \quad   \quad 
 0, \, \, 2, \, \, 3. \,    \nonumber
\end{eqnarray}
We have here an illustration of what we described in \ref{Sec:ODEformula}.1
where the third order linear differential operator $\, F_3$ reads:  
\begin{eqnarray}
F_3 \, = \, \, \, P_{sing} \cdot App(F_2)^3 \cdot App(F_3)
 \cdot {\rm D}_w^3 \,\,  + \, \, \cdots 
\end{eqnarray}
where $\, P_{sing}$ denote the ``true'' 
singularity polynomial of $\, F_3$.
We remark that the apparent polynomial of $\, F_3$ is the apparent
polynomial appearing in the product $\, F_5 \, = \, \, F_3 \cdot F_2$.
The polynomial $App(F_2)$ is the apparent polynomial of $F_2$. It appears
at the power of the order of $F_3$ for which it is a pole.
When rescaled as done in $\tilde{F}_3$ the roots of $App(F_2)$ become
apparent singularities of $\tilde{F}_3$.

Note that the formal series of the linear differential operator
$\tilde{F}_3$  are  {\em Puiseux series} around 
all the singularities except  $w=0$ and $w=\, \infty$.
These are {\it the only singular points around which
$\tilde{F}_3$ has logarithmic solutions}.
When the third order operator $\tilde{F}_3$ is rewritten in terms of the
variable $\, u \, = \, (w-w_s)^{1/2}$, where $w_s$ is any singularity
other than $w=0$ or $w=\infty$, 
 $\tilde{F}_3$ is {\em no longer singular}
at $w_s$ (in particular the ferromagnetic critical point 
$\, w\, = \, 1/4$ is {\em no longer} singular 
in the variable $\, u \, = \, \, (w-1/4)^{1/2}$).

\section{Experiment: rational reconstruction 
of the apparent polynomial in $F_3$}
\label{Sec:Experiment}
Write the  linear differential operator $F_3$ as
\begin{eqnarray}
&&F_3 \,=\,\,  {\cal P}_3(w) \,P_{app}(w^{37})\cdot {\rm D}_w^3 \,
 + {\cal P}_2(w) \, P_2(w^{51})\cdot {\rm D}_w^2\, \nonumber \\
&& \qquad \qquad 
+{\cal P}_1(w) \,P_{1}(w^{61})\cdot {\rm D}_w \, + P_0(w^{70})  \nonumber
\end{eqnarray}
where ${\cal P}_i(w)$ account for\footnote[2]{These ${\cal P}_i(w)$'s
 are different from
the ones in (\ref{49}).} the known multiplicities, and the argument
$w^{n}$ in the polynomials is used to show their respective degrees $n$.
Assume that this linear ODE has been obtained
 for many primes. We want to carry
out the rational reconstruction for each polynomial separately,
basically because the polynomials 
at the lower derivatives are harder to obtain.

As it comes from our solver, the polynomial $P_{app}$ cannot be
reconstructed with nine primes.
If we multiply all the mod prime coefficients by $2^{38}$ the rational
reconstruction will be successful with eight primes.
If we multiply by $2^{50}$ the reconstruction succeeds with six primes.
It should be noted that when the number of primes is not sufficient, the
correctly reconstructed coefficients will be those of lower degrees or
higher degrees depending on the magnitude of the scale used to multiply the
coefficients. This then calls for a scaling of the variable itself.
If we change the variable $w$ to $w/2$ and multiply all coefficients
by $2^{80}$, the rational reconstruction 
is successful with just {\em five} primes.
It is fortunate that the apparent polynomial is the easier polynomial to
reconstruct. It will be used in further checks.

How can one guess the scaling (e.g. $2^{38}$ and $2^{80}$) mentioned
above? We have found that $2^{38}$ is 
the magnitude of the lower coefficient
in ${\cal P}_3(w)$, which is an exactly known polynomial.
The scaling $2^{80}$ is around the magnitude of the lower coefficient
in $\,\, {\cal P}_3(w)\cdot P_{app}$.
More than an educated guess, we have an almost deterministic procedure
 to find the proper scaling factors to improve our rational
reconstructions. This experiment shows that the rational reconstruction
 is actually easier when the underlying physical problem is taken 
into account, leading to proper scaling factors.

\end{document}